\documentclass[aps, superscriptaddress,  twocolumn]{revtex4}
\usepackage{CJK}
\usepackage{mathtools}
\usepackage[normalem]{ulem}
\usepackage{graphicx}
\usepackage{dcolumn}
\usepackage[mathlines]{lineno}
\usepackage{hyperref}
\usepackage{bm}
\usepackage{amssymb}
\usepackage{amsmath}
\usepackage{braket}
\usepackage{color}
\usepackage{xcolor}
\usepackage{soul} 
\usepackage{graphics}
\usepackage{bbm}

\begin{document}
\title{Non-local skyrmions as topologically resilient quantum entangled states of light}

\author{Pedro Ornelas}
\affiliation{School of Physics, University of the Witwatersrand, Private Bag 3, Wits 2050, South Africa}

\author{Isaac Nape}
\affiliation{School of Physics, University of the Witwatersrand, Private Bag 3, Wits 2050, South Africa}

\author{Robert de Mello Koch}
\affiliation{School of Science, Huzhou University, Huzhou 313000, China}
\affiliation{Mandelstam Institute for Theoretical Physics, School of Physics, University of the Witwatersrand, Private Bag 3, Wits 2050, South Africa}

\author{Andrew Forbes}
\affiliation{School of Physics, University of the Witwatersrand, Private Bag 3, Wits 2050, South Africa}

\begin{abstract}
\noindent In the early 1960s, inspired by developing notions of topological structure, Tony Skyrme suggested that sub-atomic particles be described as natural excitations of a single quantum field.  Although never adopted for its intended purpose, the notion of a skyrmion as a topologically stable field configuration has proven highly versatile, finding application in condensed matter physics, acoustics and more recently optics, but all realised as localised fields and particles. Here we report the first non-local quantum entangled state with a non-trivial topology that is skyrmionic in nature, even though each individual photon has no salient topological structure. 
We demonstrate how the topology of the quantum wavefunction makes such quantum states robust to entanglement decay, remaining intact until the entanglement itself vanishes.  Our work points to a nascent connection between entanglement classes and topology, holding exciting promise for the creation and preservation of quantum information by topologically structured quantum states that persist even when entanglement is fragile.

\end{abstract}
\maketitle


\noindent In the early 1960s Tony Skyrme proposed a non-linear meson field theory to describe sub-atomic particles as excitations of a single fundamental field, the pion \cite{skyrme1962unified,ZAHED19861, Naya2018Skyrmions,eisenberg1981nucleon}.  To accomplish this he used the inherent mathematical structure of basic pion theory to postulate topologically non-trivial pion field configurations, now called skyrmions.  A skyrmion is a topologically stable field configuration, characterized by an integer topological invariant, the skyrme number. A skyrmion cannot be smoothly deformed into another field configuration with a different skyrme number. The generality of this definition and its associated topological conserved quantity allowed the notion to be extended beyond its initial intent \cite{shen2022topological,he2022towards}.  In particular, skyrmions have been instrumental in advances in magnetism and spintronics, where their topological stability makes them ideal candidates for information storage and transfer \cite{yu2010real,fert2013skyrmions,fert2017magnetic,Nagaosa2013,Zhang2020,LimaFernandes2022,zheng2022skyrmion}. Here the quantum properties of these localised magnetic textures have been theoretically studied \cite{psaroudaki2017quantum,psaroudaki2022skyrmion,lohani2019quantum,douccot2008entanglement,froehlich1990quantum,siegl2022controlled} and suggested as a basis for quantum information processing \cite{psaroudaki2021skyrmion}, where their macroscopic quantum
tunneling and energy-level quantization are indicative of
quantum behavior, with some quantum dynamics already observed in magnetic systems \cite{zhou2020solids}.

Beyond magnetism, skyrmions have been revisited in the context of nuclear physics to resolve long standing debates on the nucleon-nucleon spin-orbit potential \cite{halcrow2020attractive} and have been observed in atomic matter \cite{leslie2009creation}, chiral liquid crystals \cite{ackerman2015self} and in acoustics \cite{ge2021observation}. Optical realisations have only recently been explored, including observations in evanescent waves \cite{tsesses2018evanescent}, in focused orbital angular momentum (OAM) beams at sub-wavelength scales \cite{du2019deep}, in certain classes of full Poincar\'{e} beams \cite{gao2020paraxial,kuratsuji2021evolution,shen2022generation} and even as toroidal pulses \cite{shen2021supertoroidal}.  Tracking the trajectories of individual psuedospin states with propagation exposes a mapping to the 4-dimensional hypersphere, realizing the skyrme field as a Hopf fibration \cite{sugic2021particle}.   

\begin{figure*}[t]
\includegraphics[width=0.9\linewidth]{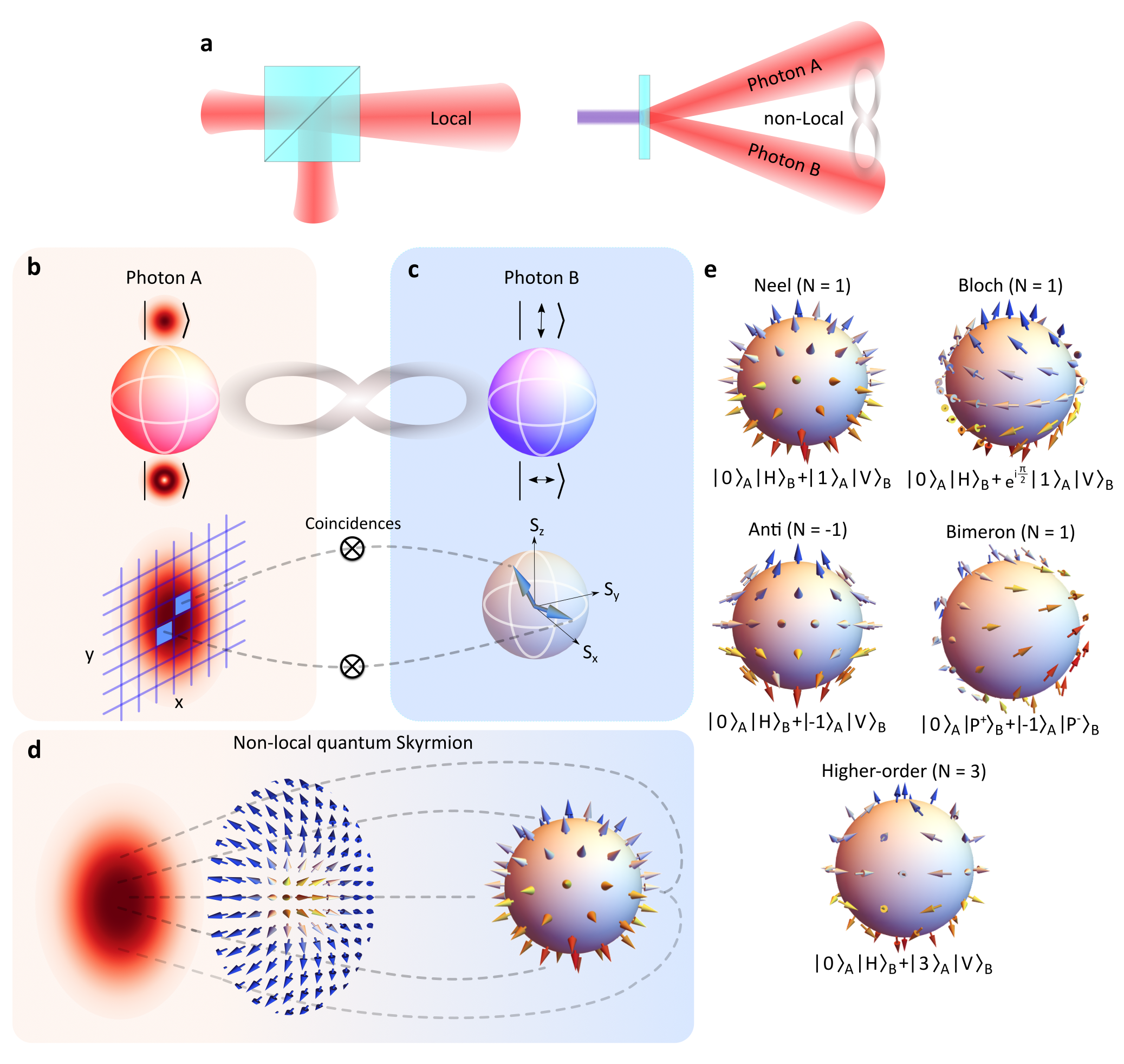}
\caption{\textbf{Non-local quantum skyrmions.} \textbf{a}, A typical local field configuration giving rise to a skyrmion can be produced by superimposed, orthogonally polarized spatial modes. The non-local quantum skyrmion exists as a shared property of spatially separated photons A and B. The non-local skyrmionic configuration is formed in the entanglement between the two photons, with \textbf{b}, photon A carrying the spatial degree of freedom (DoF) expressed on an OAM Bloch Sphere, and \textbf{c}, photon B carrying the polarization DoF expressed on the polarisation Bloch Sphere. A spatial measurement on photon A collapses photon B into some definite polarization state, linking space to polarisation non-locally by a coincidence measurement (bottom panel). A complete spatial mapping of photon A reveals all polarisation possibilities in photon B. \textbf{d}, The entanglement between the photons is now skyrmionic, as denoted by the vectorial arrows in space, which can be mapped stereographically to an abstract sphere that holds spatial information (coverage) and the skyrmionic state (arrows and location). \textbf{e}, Full control over the quantum wavefunction allows for the manipulation of the texture and precise topology of the non-local skyrmion thereby giving access to a plethora topological structures.}

\label{fig:ConceptFig}
\end{figure*}

These impressive advances are all local realisations of skyrmions as particles and fields, without any essential role for entanglement, as illustrated in Figure~\ref{fig:ConceptFig}a. Creating skyrmions as quantum entangled states with non-local quantum correlations would allow potent applications of spatially structured topologies to quantum topological photonics \cite{yan2021quantum}.  Despite being very much in its infancy, the merging of topological and quantum structure holds great promise for information robustness even in non-ideal quantum systems, including stable quantum emitters \cite{mehrabad2020chiral,dai2022topologically,mittal2018topological} and robust transport of quantum states through quantum circuits \cite{barik2018topological,blanco2018topological}. 

Here we report the first non-local quantum entangled state with a non-trivial topology that is skyrmionic in nature.  Intriguingly, the non-trivial topological structure does not exist in the properties of the individual (local) photons in the two-photon entangled state, but rather it emerges from the (non-local) entanglement between them.  We show that the topology of the quantum wavefunction makes this skyrmionic state robust to entanglement decay, remaining intact until the entanglement itself vanishes, and here we report the first demonstration of this effect.  Our approach allows the topology of these quantum wavefunctions to be fully controlled, which we outline theoretically and demonstrate experimentally across a wide variety of skyrmion types, and allowing us to distinguish entanglement classes according to their quantum topology. Our work leverages on topological photonics and quantum state engineering, offering a promising avenue for preservation of quantum information by topologically engineered quantum states that persist even when entanglement is fragile.  


\vspace{0.5cm}
\noindent \textbf{Concept.}  An optical skyrmion can be represented as a topologically protected spin textured field, where every point on the Poincar\'{e} Sphere (${\cal S}^2$) is in correspondence with a point in the 2D transverse spatial plane (${\cal R}^2$).  The optical skyrmion is a mapping from the 2D transverse spatial plane to the space of polarisation states, i.e., a mapping ${\cal R}^2 \rightarrow {\cal S}^2$. We create this by entanglement, through engineering two entangled photons (A and B) to occupy the quantum state $\ket{\Psi} = \frac{1}{\sqrt{2}} \left ( \ket{\ell_1}_A \ket{H}_B + e^{i\gamma}  \ket{\ell_2}_A \ket{V}_B \right )$, where $\ell_1$ and $\ell_2$ denote OAM of $\ell_1 \hbar$ and $\ell_2 \hbar$ per photon, respectively, $H,V$ are the orthogonal horizontally and vertically polarised states, while $\gamma$ allows for a rotation of the state vector.  The reduced state of photon A is an incoherent mixture of OAM, while that of photon B is unpolarised, so that \textit{individually} neither has any salient topological structure. Quantum correlations between the two photons imply a rather different picture: the collapsed state of one photon is determined by the measurement choice on the other, so that a joint measurement on both reveals \textit{non-local} quantum topological structure.  Intriguingly, each localised position of photon A in real space $\mathcal{R}^2$ is associated with a polarisation state, occupied by its entangled twin (photon B) and parameterised on a sphere, $\mathcal{S}^2$. We expand on this mathematically later in the context of a wider meaning. Consequently photon A holds the possibility for every spatial position, shown in Figure~\ref{fig:ConceptFig}b as both a spatial mode Bloch Sphere (top) and the corresponding spatial probability distribution (bottom). Photon B meanwhile holds the possibility for every polarisation state, shown in Figure~\ref{fig:ConceptFig}c as both a polarisation Bloch Sphere (top) and the corresponding Poincar\'{e} Sphere (bottom).  A spatial measurement on photon A collapses photon B into a particular polarization state, producing a mapping from space to the Poincar\'{e} Sphere, revealed by joint measurements in coincidence, so that we have indeed obtained the mapping ${\cal R}^2 \rightarrow {\cal S}^2$ as desired.  While the topological structure of each individual photon is always trivial and with zero skyrme number, the non-local skyrme number (we call it the quantum skyrme number) denoted by $N$, can be tailored as desired, with the skyrmion existing in the entanglement between the photons themselves. This non-local quantum entangled skyrmion is depicted graphically in Figure~\ref{fig:ConceptFig}d.  The skyrmion topology is shown as vectorial arrows in space, the position and direction of which are derived from the joint state of photons A and B. Photon A also contributes the probability of detection, spatially varying, shown as a false colour density plot. This information can be visualised holistically by a stereographic mapping to an abstract sphere whose spatial coverage is given by photon A and the skyrmion topology (position and directions of the arrows) by the joint state of photons A and B. Furthermore, with complete control over the quantum wavefunction the topology and texture can be modified at will, examples of which are given in Figure~\ref{fig:ConceptFig}e.

\begin{figure*}[t]
\includegraphics[width=\linewidth]{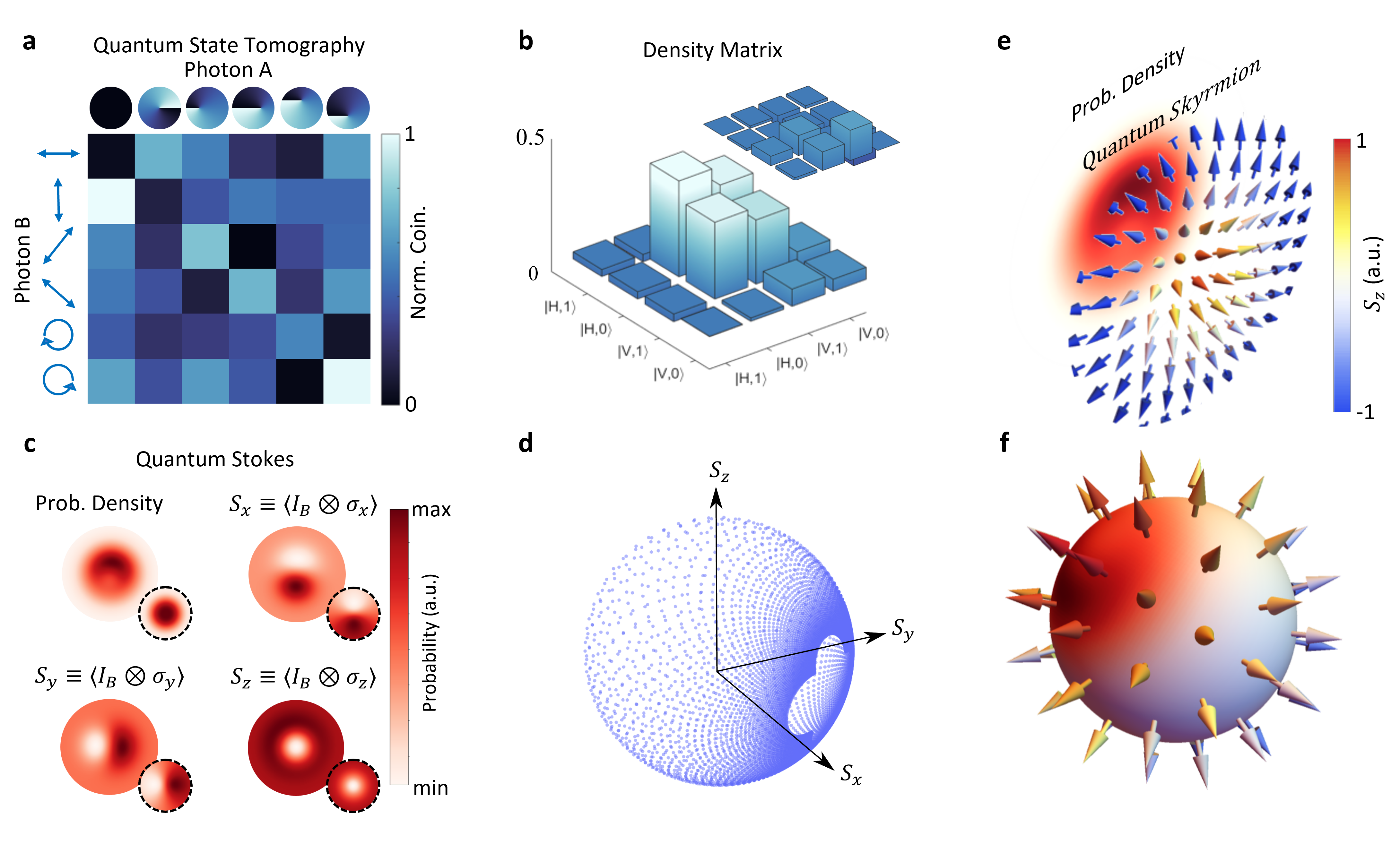}
\caption{\textbf{Experimental quantum skyrmion.} \textbf{a}, A quantum state tomography was performed on the selected entangled state, $|\Psi\rangle = \frac{1}{\sqrt{2}}\left(|1\rangle_A|H\rangle_B + |0\rangle_A|V\rangle_B\right)$ by spatial measurements on photon A (columns) and polarisation measurements on photon B (rows), with coincidences collected for all outcomes shown as false colour from low (black) to high (white). \textbf{b}, The real and imaginary (inset) parts of the density matrix for the reconstructed state, indicative of the desired state. \textbf{c}, Measured quantum Stokes projections directly from the tomography. \textbf{d}, Reconstruction of the coverage on the Poincar\'{e} Sphere, with experimental data shown as blue dots. \textbf{e}, The experimental probability density of finding a photon in space, given by the spatial structure of photon A, and the experimentally reconstructed skyrmion as a spin textured field with a defined topology. \textbf{f}, A stereographic mapping of the experimental data showing spatial information (coverage) and probability of detection (colour), with the skyrmionic state depicted by the direction of the arrows and their location, a direct outcome from the joint state of photons A and B.}
\label{fig:ExpFig}
\end{figure*}
\vspace{0.5cm}
\noindent \textbf{Creating and detecting quantum skyrmions.} To verify this concept, we prepared our entangled state from an initial spontaneous parametric downconversion (SPDC) source and performed a unitary operation on the OAM of Photon A while performing a spatial-to-polarization (SPC) conversion, exchanging the spatial information of photon B for polarization information, with full experimental details given in the Supplementary Information (SI).  This altered the typical SPDC OAM spiral spectrum to the desired asymmetric spiral bandwidth for a skyrmionic two-photon quantum wavefunction.  The non-local two-photon wavefunction was then analysed by projective measurements with a spatial light modulator (SLM) on photon A and polarisation projections on photon B, measured in coincidence.  The results from this quantum state tomography (QST) are shown in Figure~\ref{fig:ExpFig}a for an example state with $\ell_1 = 1$ and $\ell_2 = 0$ and $\gamma = 0$, where the detected coincidences for each projective measurement has been normalized against the maximum coincidences detected. The top-left $2 \!\! \times \!\! 2$ matrix partially depicts the behaviour of the generated state, that is that when projecting onto $\ket{\ell}=\ket{0}$ for photon A, we only detect coincidences for vertically polarized photon B and similarly when projecting onto $\ket{\ell}=\ket{1}$ we only detect coincidences for horizontally polarized photon B. This is consistent with our expected state behaviour, a measurement of photon A revealing $\ell = 0(1)$ collapses photon B into the state $\ket{V} (\ket{H})$. Further projections are performed to build an over-complete QST (further details on QST are given in the SI) which enabled the reconstruction of the density matrix for our skyrmionic quantum state, shown in Figure~\ref{fig:ExpFig}b. The real and imaginary (inset) parts shown contain all the information of our quantum state. When compared with the expected pure theoretical state, the experimentally generated state revealed a fidelity of $F = 95.0\%$. Additionally, the QST contains the information necessary to reconstruct the non-local Quantum Stokes parameters for the state, shown in Figure~\ref{fig:ExpFig}c after local normalization, with the theoretical Quantum Stokes parameters given as insets. The probability density is given by $\langle I_A \otimes I_B \rangle$ and depicts the probability of finding photon A in a particular position and photon B in the corresponding polarization state. Furthermore, the remaining Quantum Stokes parameters are calculated as the expectation values of the Pauli matrices with further detail on how this information is extracted from QST given in the SI. The Quantum Stokes parameters reveal a mapping of each point in space to a position on the Poincar\'e sphere such that our complete set of spatial measurements on photon A reveals a collapse of photon B into 97.3\% of all possible polarization states parametrized by the Poincar\'e sphere. This coverage is shown in Figure~\ref{fig:ExpFig}d. In Figure~\ref{fig:ExpFig}e the true skyrmionic nature of the generated biphoton state is depicted. The probability density and spatial information is extracted by measurement of photon A and in coincidence the polarization information is extracted from measurement of photon B, revealing a Neel-type configuration. A stereographic projection of the configuration shown in Figure~\ref{fig:ExpFig}e yields the hedgehog-like polarization texture shown in Figure~\ref{fig:ExpFig}f of the entangled state on the surface of a sphere. An experimental quantum skyrme number of $N_\text{exp} = 0.972$ which is very close to the theoretical value of $N_\text{th} = 1$, for the first realisation of a non-local quantum entangled skyrmion.

\begin{figure*}[t!]
\includegraphics[width=0.75\linewidth]{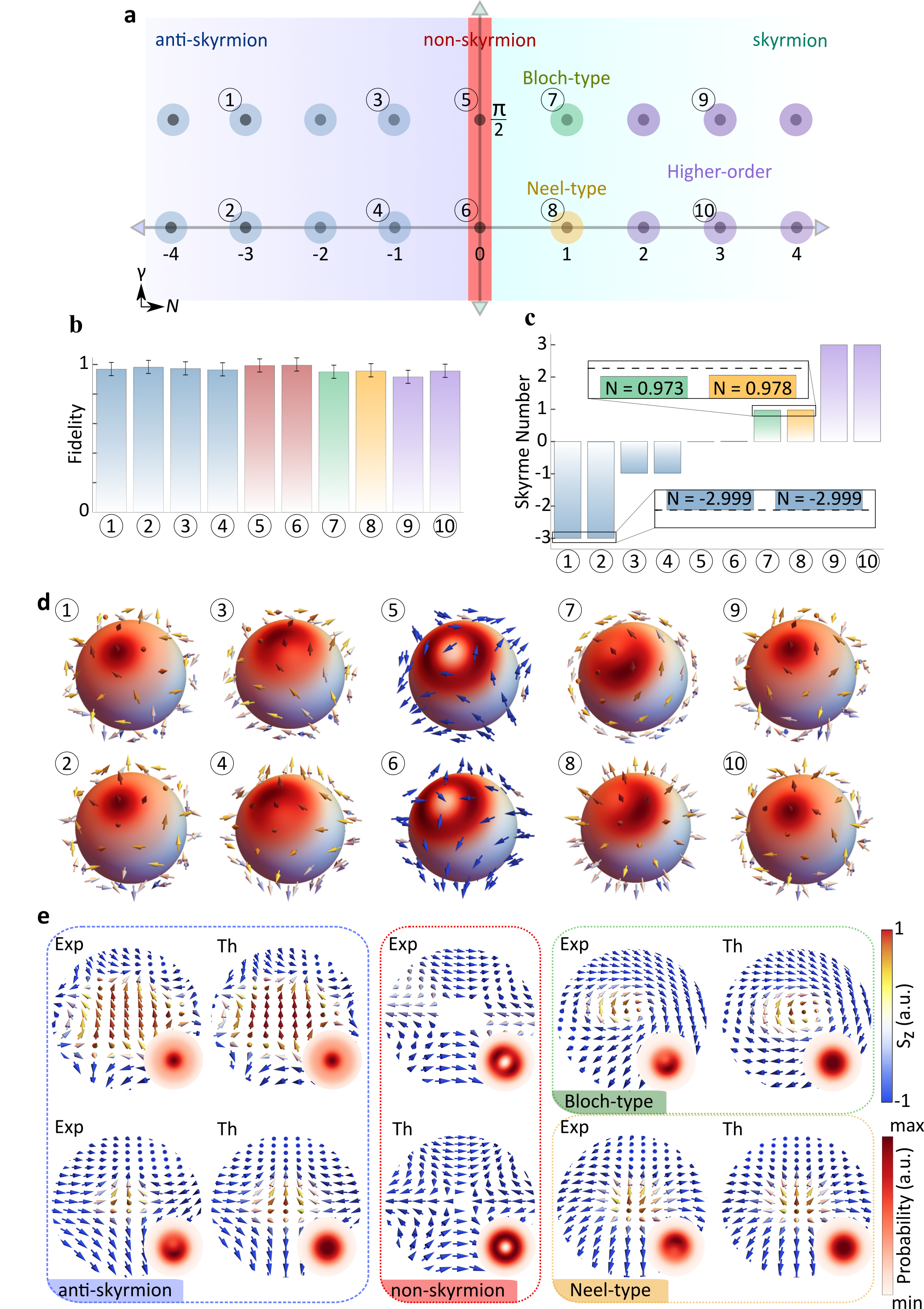}
\caption{\textbf{Traversing the quantum skyrmionic landscape.} \textbf{a}, Varying textures (down a column) and skyrme numbers (across a row) were created by controlling $\gamma$ and the OAM difference ($\Delta \ell$) in the quantum state to indirectly control $N$. Example states are shown graphically and labelled with numbers 1 through 10, traversing the skyrmionic landscape from anti-skyrmions ($N < 0$) through non-skyrmionic states ($N = 0$) to Bloch-type ($N=1,\gamma = \pi/2$), Neel-type ($N=1,\gamma = 0$) and higher-order skyrmions $N>1$. \textbf{b}, Experimental fidelities for the example states when compared to theoretical maximally entangled pure states. \textbf{c}, Experimental skyrme numbers for example states. \textbf{d}, Examples of experimental stereographic projections. \textbf{e}, Example skyrmion topologies with varying textures shown as experimental (Exp) and theoretical (Th) reconstructions.  The insets show the probability of detection as measured (Exp) and calculated (Th) from photon A.}
\label{fig:landscapeFig}
\end{figure*}

Next, we traverse the skyrmionic quantum landscape, altering the texture and topology by control of $\gamma$ and $N$, respectively.  We illustrate this in Figure \ref{fig:landscapeFig}a, moving from anti-skyrmions ($N < 0$) through non-skyrmionic states ($N = 0$) to $N=1$ examples of Bloch-type ($\gamma = \pi/2$) and Neel-type ($\gamma = 0$) and lastly to higher-order skyrmions ($N>1$), all faithfully produced with high fidelity when compared to maximally entangled states, as shown in Figure \ref{fig:landscapeFig}b and measured skyrme numbers ($N_\text{exp}$), as shown in Figure \ref{fig:landscapeFig}c in excellent agreement with theoretically predicted integer values. See SI for exact fidelity and skyrme number values for the generated states referenced in Figure \ref{fig:landscapeFig}.  In Figure \ref{fig:landscapeFig}d we show the experimentally inferred topology of the generated states as stereographic projections onto the surface of $S^2$, while in \ref{fig:landscapeFig}e we show selected experimental and theoretical topologies (with the experimental and theoretical probability densities of each state given as insets), in excellent agreement.  Notice that when $\ell_2 = -\ell_1$ i.e., opposite twists to the OAM components, as would be created in typical spin-orbit hybrid entanglement experiments \cite{forbes2019quantum}, no quantum skyrmion is produced and $N = 0$. For $N=-1$, we see the characteristic hyperbolic texture embedded into the wavefunction, and for the Neel and Bloch type skyrmions we see the characteristic hedgehog and spiral textures.  Full experimental data sets are given in the SI.

\begin{figure*}[t!]
\includegraphics[width=0.8\linewidth]{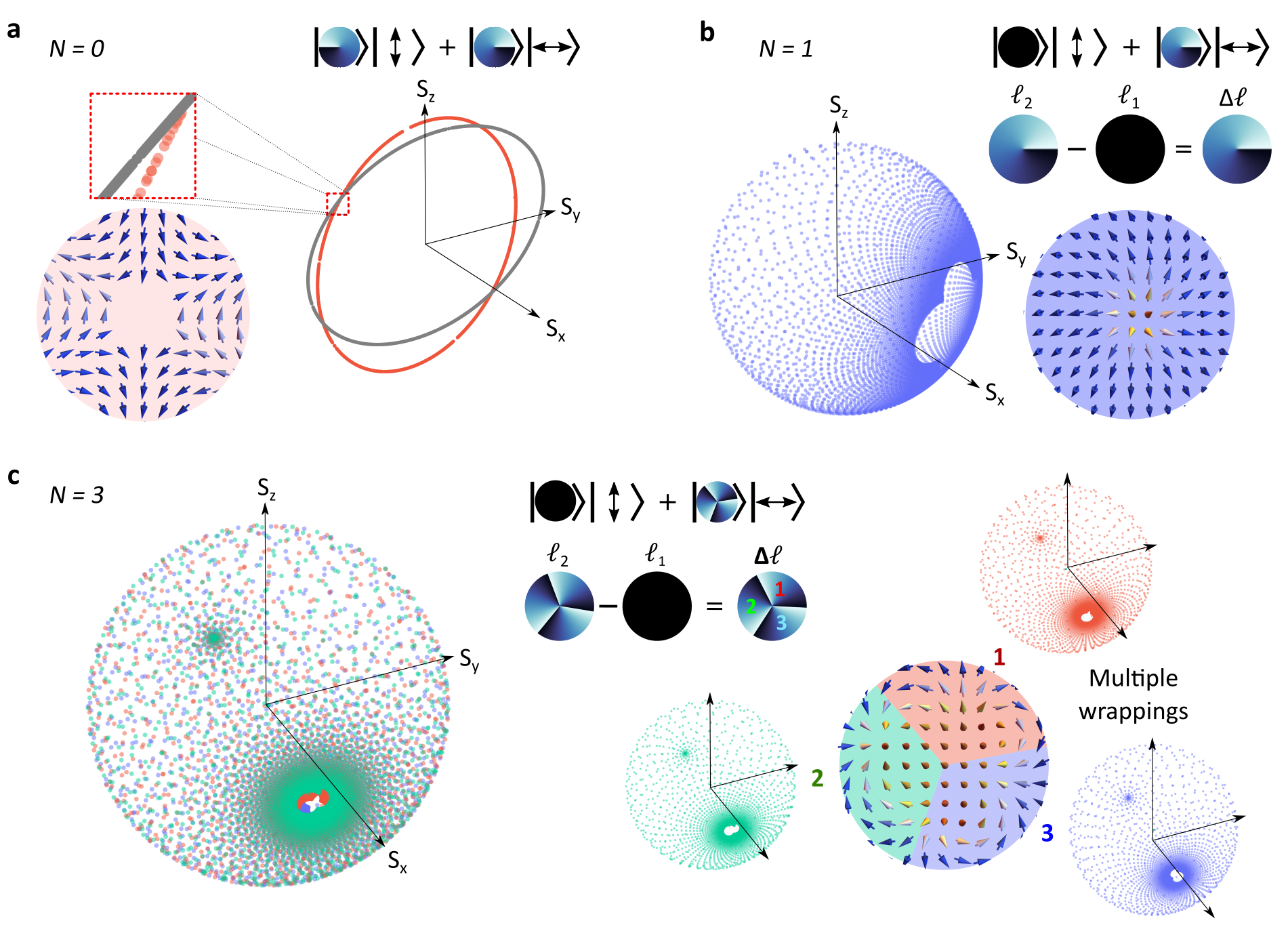}
\caption{\textbf{Topology of quantum entangled states.} Bi-photon entangled states can be classified according to their topology. \textbf{a}, Entangled States where $|\ell_1| = |\ell_2|$ have $N=0$ and map to rings on $\mathcal{S}^2$ with close to zero coverage and are topologically equivalent. Experimental results (dots) together with the perturbation-free prediction (solid line) are shown in the inset. States with $|\ell_1| \neq |\ell_2|$ necessarily have $N \neq 0$ and map to the entirety of $\mathcal{S}^2$, through $N$ wrappings. This is shown for (\textbf{b}, $N=1$ and \textbf{c}, $N=3$, with experimental data is shown as dots on the spheres. In \textbf{b} the coverage is close to 100\% for $N=1$. \textbf{c}, For $N=3$, the non-local field has three segments, each with a full coverage of $\mathcal{S}^2$, shown as red, green and blue experimental dots. The entire state maps to $\mathcal{S}^2$ with three windings for a coverage of $299\%$, close to 300\%, illustrated in the composite sphere (right) with all three experimental data sets.}
\label{fig:classesFig}
\end{figure*}
\vspace{0.5cm}

\noindent \textbf{Topology and entanglement.} A convenient basis of states in which to expand the wavefunction of photon A is provided by the position space states $\ket{\mathbf{r}_A}$. Using the explicit expression $\ket{\ell} = \int_{ \mathcal{R}^2 } |\text{LG}_\ell \left(  \mathbf{r_A} \right)| e^{ i \ell \phi_{A}} \ket{\mathbf{r}_A} d^2 r_A$ for the OAM eigenstates, we write our entangled photon state as
\begin{align}
 \ket{\Psi} = \int_{ \mathcal{R}^2 }  \ket{\mathbf{r}_A} \left( a(\mathbf{r}_A) \ket{H}_B + b(\mathbf{r}_A) e^{ i\Theta(\phi_A)}\ket{V}_B \right) d^2 r_A,  \label{Eq:SpatialquantumSkyrmion}
\end{align}
\noindent 
where $a(\mathbf{r}_A) \equiv |\text{LG}_{\ell_{1}} \left( \mathbf{r}_A \right)|$, $b(\mathbf{r}_A) \equiv |\text{LG}_{\ell_{2}} \left( \mathbf{r}_A \right)|$, $\Theta(\phi_A) = \Delta\ell\phi_A + \gamma$ and $\Delta\ell = \ell_2-\ell_1$. The coefficient of $\ket{\mathbf{r}_A}$ above defines a state for photon $B$
\begin{align}
 \ket{\psi_{B|A}} = \cos(\theta(\mathbf{r}_A)) \ket{H}_B +  \sin(\theta(\mathbf{r}_A)) e^{ i\Theta(\phi_A)}\ket{V}_B ,  \label{Eq:reducedstate}
\end{align}
where the $\theta(\mathbf{r}_A)$ dependence simply reflects the fact that we have normalized the state. The polarization state space for photon $B$ is a two dimensional sphere $\mathcal{S}^2$. Using a stereographic projection, the original position state space $\mathcal{R}^2$ can also be identified as an $\mathcal{S}^2$, that is, we can map it to a sphere. In this way, by associating $\ket{\mathbf{r}_A}$ and $ \ket{\psi_{B|A}}$, the entangled photon wavefunction defines a map from a sphere to a sphere: from the position state space $\mathcal{S}^2$ (associated to photon $A$) to the polarization state space $\mathcal{S}^2$ (associated to photon $B$).

Every map from $\mathcal{S}^2$ to $\mathcal{S}^2$ has an integer degree, $N$, measuring how many times the first sphere wraps the second. This degree is a topological invariant: it is unchanged by continuous deformation of the map. This rich topological structure of the maps from spheres to spheres is the source of the topological structure in our quantum two-photon wavefunctions. In the current context, the degree $N$ is the quantum skyrme number. An important corollary of the existence of a topological invariant is that the space of continuous maps is partitioned into equivalence classes \cite{hatcher2002algebraic}. Any two maps in the same class can be continuously deformed into each other, while this is not possible for maps in distinct classes \cite{hatcher2002algebraic}. The connection we have established between entangled photon wavefunctions and maps from $\mathcal{S}^2$ to $\mathcal{S}^2$ implies that the space of all possible wavefunctions is itself partitioned into equivalence classes with each class labeled by its quantum skyrme number.

To illustrate this, we show examples of experimentally measured mappings with various winding numbers in Figure~\ref{fig:classesFig}. All entangled states with $N=0$ map to rings on $\mathcal{S}^2$ and hence are topologically equivalent, with the results in Figure~\ref{fig:classesFig}a confirming the mapping. This topology groups conventional hybrid entangled states, as is usually created by geometric phase approaches \cite{stav2018quantum}, into a single class.  Note the deviation between the perturbation-free prediction (solid line in the inset) and the experimental points, which we return to shortly in the context of robustness.  States with $N \neq 0$ wrap the entire $\mathcal{S}^2$ space $N$ times, as shown in Figure~\ref{fig:classesFig}b and Figure~\ref{fig:classesFig}c for $N=1$  and $N=3$, respectively. The experimentally measured coverage of approximately 97\% ($N=1$) and 299\% ($N=3$) come close to the theoretical values of 100\% and 300\%, respectively. The small deviation can be seen at the pole of of the mapped sphere, a natural artefact of mapping from $\mathcal{R}^2$ to $\mathcal{S}^2$, where data at spatially infinite distances cannot be recovered. Furthermore we note that for any entanglement class, the spatial state space may be partitioned into $N$ segments, each of which will be found to wrap the entire $S^2$ state space, as shown in Figure~\ref{fig:classesFig}c. 

Having identified the origin of the topological structure in the two-photon quantum state, we now offer a connection between topology and entanglement. A non-trivial topology is not a property of either of the $\mathcal{S}^2$ spaces on their own, but rather it is a shared emergent property originating in the non-trivial global structure of the map between the two spaces. Entanglement itself is not a property of the state of either particle on its own, but rather it too is a shared emergent property. There is a quantitative relationship between the two: it is only the topologically trivial $N=0$ class of two photon wavefunctions that can have vanishing entanglement. Wavefunctions belonging to any other class are always entangled. This argument is supported by noting that since for the trivial $N=0$ topology do not wrap the $\mathcal{S}^2$ we map to, it is possible to continuously deform the map so that all points on the first sphere map to a single point on the second sphere. This map corresponds to a wavefunction for which every position state $\ket{\mathbf{r}_A}$ in Equation \ref{Eq:SpatialquantumSkyrmion} multiplies the same state from the polarization state space. This product state is not entangled. For {\it any} other winding number $N\ne 0$, the coefficient of different position space states range $N$ times over the complete polarization state space, so the corresponding state is entangled. The inescapable conclusion is that non-trivial topology $N\ne 0$ implies non-zero entanglement. On the other hand, trivial topology does {\it not} imply that the corresponding wavefunction has vanishing entanglement. Indeed, wavefunctions corresponding to maps that partially cover the polarization $\mathcal{S}^2$ are entangled, but still topologically trivial with $N=0$. The point however, is that these maps can be continuously deformed to a map which corresponds to the zero entanglement wavefunction. For any other $N\ne0$ class, it is not possible to reach the map of a zero entanglement wavefunction by continuous deformations. In this sense the topology of the wavefunction partitions the space of all possible wavefunction into entanglement classes. 

\begin{figure*}[t]
\includegraphics[width=\linewidth]{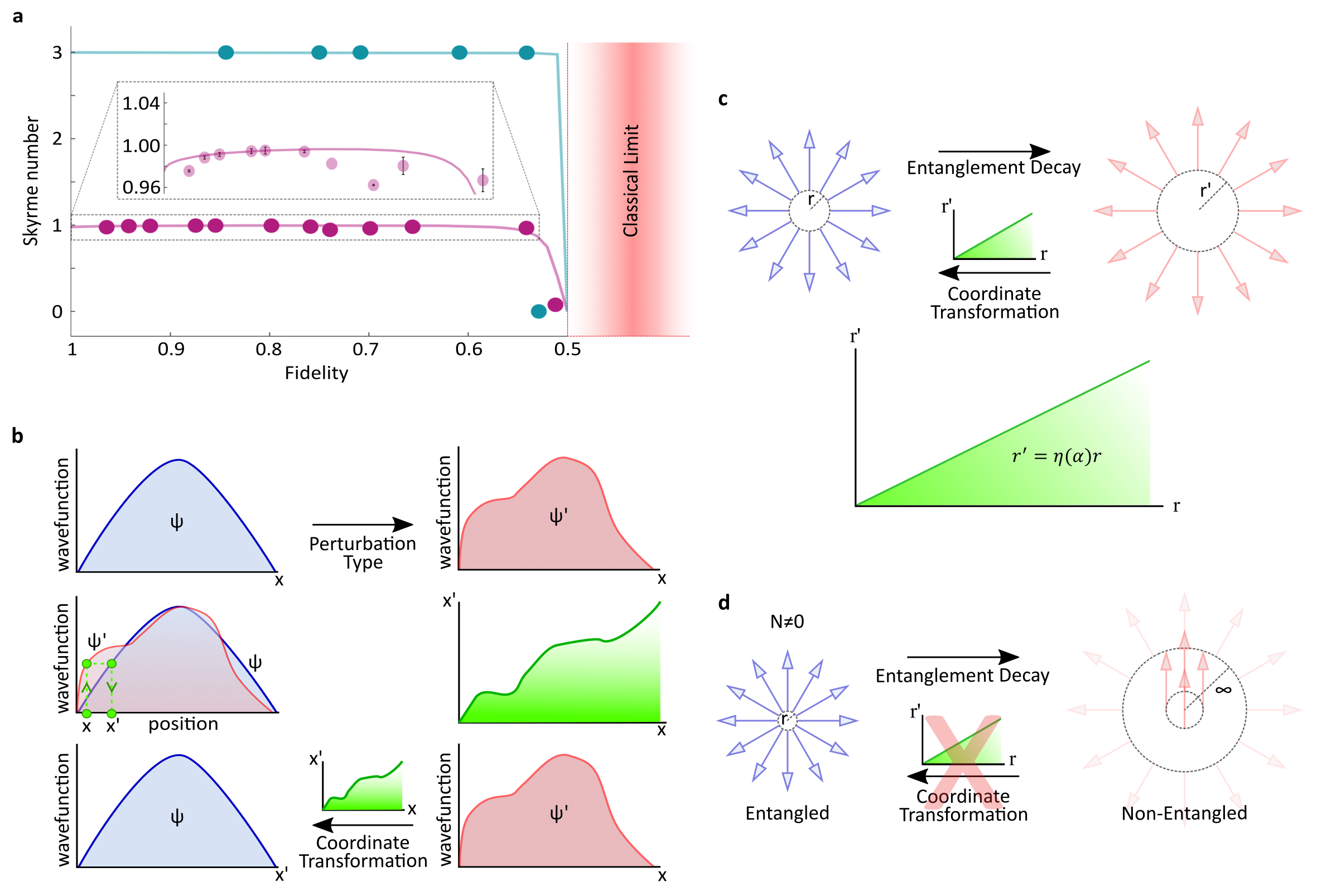}
\caption{\textbf{Quantum topological invariance.} \textbf{a}, Measured quantum skyrme number ($N$) as a function of state fidelity ($F$) against a maximally entangled state.  Here $F$ decreased by a deliberately induced decay in entanglement, from fully entangled ($F = 1$) to no entanglement ($F = 0.5$). Results are shown for initial $N = 1$ and $N = 3$, with the invariance of $N$ clearly revealing that the topology remains intact in the presence of decaying entanglement. Experimental data is shown as dots and theory as solid lines, in excellent agreement with one another. The errors for calculated skyrme number $N = 1$ is shown as an enlarged inset. \textbf{b}, The coordinate transformation defined by the original and distorted wavefunctions. To obtain the mappping $x \rightarrow x'$ we follow point $x$ vertically up to $\psi'$, then traverse horizontally from $\psi'$ to $\psi$ and finally project down from $\psi$ to $x'$. In this way, the distorted wavefunction is mapped back to the original wavefunction through a coordinate transformation. \textbf{c}, The entanglement decay operation results in a simple smooth deformation of the wavefunction which can be seen physically as a scaling of the radial position of any set of polarization vectors that lie at some radial position. A simple linear radial coordinate transformation therefore returns the wavefunction to its original state. \textbf{d}, When the entanglement has vanished so has the non-trivial topology, thus there is no coordinate transformation that can be realised to return the state to its previous form.}
\label{fig:stable}
\end{figure*}

\vspace{0.5cm}
\noindent \textbf{Topological resilience.} Our connection between topology and entanglement suggests that (1) it is not possible to have no entanglement and a non-trivial topology ($N \neq 0$), (2) it is possible to have a trivial ($N=0$) topology and no entanglement, while (3) the local deformation preserving topology suggests that so long as the entanglement persists the topology is robust  - a quantum form of topological resilience.  The results reported in Figure ~\ref{fig:stable}a support these statements.  We introduced an operation applied to the wavefunction that resulted in entanglement decay (see Methods and SI) from a maximally entangled state to no entanglement, all the while monitoring the topology through the quantum skyrme number, $N$. The results for $N = 1$ and $N=3$ reveal a new form of topological resilience, with the quantum skyrme number intact in the presence of decaying entanglement, satisfying (1) and (3), falling off to zero only when the entanglement itself vanishes, satisfying (2). To the best of our knowledge, this is the first observation of robustness to entanglement decay derived from the topology of the entanglement between particles, which holds exciting promise for topologically protected quantum information processing.

We now sketch a simple argument explaining how this occurs for quantum wavefunctions which we believe is not to be found elsewhere. As a first step, can we understand what we gain by coding information into topology? Topology is a systematic approach to characterizing those quantities that are insensitive to smooth deformation.  Our observed topological resilience of the skyrmionic wavefunction can be explained by considering the entanglement decay operation as a smooth deformation of the state which in our problem is the statement that the quantum skyrmion number is unchanged by a change of coordinates, $(x,y) \rightarrow (x',y')$, for photon A (see SI for a proof).  After a short derivation (see SI for details) we find that indeed
\begin{equation}
    N = \frac{1}{4\pi}\int\limits_{-\infty}^{\infty}\!\int\limits_{-\infty}^{\infty} \Sigma_z (x,y)  dx dy
      = \frac{1}{4\pi}\int\limits_{-\infty}^{\infty}\!\int\limits_{-\infty}^{\infty} \Sigma_z (x',y')  dx' dy' 
    \label{eqn:TopInvSkyNum}\,
\end{equation}
where
\begin{equation}
    \Sigma_z (x,y)= \frac{1}{2} \epsilon_{pqr} S_p \frac{\partial S_q}{\partial x} \frac{\partial S_r}{\partial y}\,
\end{equation}
and $x',y'$ are both arbitrary, smooth functions of $x,y$. This remarkable property is proved by noting that in moving from the first to second equality in Equation \ref{eqn:TopInvSkyNum}, $\Sigma_z$ picks up a factor of the inverse Jacobian while the measure picks up a factor of the Jacobian, so the product is invariant (see the SI for a full proof). This is the complete content of the statement that the quantum skyrmion number is invariant under smooth deformations. 
As a second step we must explain how this leads to resilience. It is simplest to phrase the discussion in a single spatial dimension, which we do with the aid of Figure~\ref{fig:stable}b. Consider a perturbation that modifies a wavefunction $\psi(x)$ into $\psi'(x)$. If the perturbation is modest, the equality $\psi'(x)=\psi(x')$ defines a change of coordinates from $x$ to a new coordinate $x'$, as seen graphically in Figures~\ref{fig:stable}b. The topological invariant, our quantum skyrme number, is unchanged by this coordinate transformation, so we are free to perform it. The transformation replaces $\psi'(x)\to\psi(x')$, that is, a simple coordinate change allows one to map from the distorted wavefunction to the original wavefunction.  The return to our undistorted wavefunction by this simple coordinate transformation completes the demonstration of topological resilience. This argument is a mathematical demonstration of the fact that the topology is a global property of the wavefunction, insensitive to the local changes induced by perturbations.

Returning to the experimental demonstration of the topological resilience to entanglement decay Figure~\ref{fig:stable}a, we can now discuss this in terms of the arguments provided above. We can indeed observe the entanglement decay operation as a smooth deformation of the entangled state as it manifests as a radial position coordinate scaling for polarization state vectors which lie at some $r$, as shown in Figure~\ref{fig:stable}c, so that a simple linear coordinate transformation of the form $r \to r'$ where $r' = \eta(\alpha) r$ yields the original entangled state (Full details of the coordinate transformation derivation is given in the SI). We see from Figure~\ref{fig:stable}a that as we asymptotically approach vanishing entanglement, so the topology abruptly changes from non-trivial ($N \neq 0$) to trivial ($N = 0$), emphasising that the skyrmionic topology is a \textit{shared} property of the entangled photons, which necessarily vanishes once there is no entanglement.

Because perturbing the wavefunction corresponds to deforming the map, information about the entanglement class of the wavefunction is naturally resilient to such a perturbation. Further, any topologically non-trivial ($N\ne0$) class enjoys this resilience, while the topologically trivial class does not.  An experimental demonstration of this can be visualised by the mappings in Figure \ref{fig:classesFig}: since the topologically trivial class is mapped to a ring, as shown in Figure \ref{fig:classesFig}a, its orientation can easily be discerned.  Indeed, in Figure \ref{fig:classesFig}a we can see the deviation from the undistorted prediction of the ring (solid line) and the experimental reconstruction (data points), the latter always with some inherent perturbation. Conversely, one cannot discern an orientation change due to perturbation when the entire sphere is mapped, as is the case for all non-trivial classes with examples shown in Figure \ref{fig:classesFig}b and Figure \ref{fig:classesFig}c: full (or more) coverage always looks the same regardless of your perspective.  

\subsection*{Discussion and Conclusion}

We have reported the first realisation of skyrmionic fields with non-local quantum correlations, which have hitherto been elusive. The topologically non-trivial two-photon entangled state has a skyrme number that can be any non-zero integer, and this topological structure is not derived from either local photon state, but rather, from the entanglement between them. We demonstrated experimental control over the topology of our two-photon entangled states showing exotic topologies ($N \in \{-3,-1,0,1,3\}$) and textures (anti-, Neel- and Bloch types). We have outlined the classification of these entangled states according to their topology, from trivial classes that map to rings on $\mathcal{S}^2$, to non-trivial classes with $N$ wrappings of $\mathcal{S}^2$, shown up to $N=3$. We have shown that such quantum textures imbue the state with the topological resilience, preserving topology until the entanglement itself vanishes and establishing that one may identify an entangled state's topology even when entanglement is fragile. Our results should inspire new directions in engineering topological quantum states, for instance for a quantum version of vectorial holography \cite{song2022vectorial}, and could benefit from resonant metasurfaces for compact sources of such states \cite{zheludev2012metamaterials}.  While fuelling new research avenues, we believe that our work holds exciting prospects for information robustness by topology even in non-ideal quantum systems, and may be harnessed for resilient quantum information processing and communication in quantum networks that use entanglement as a resource.

\newpage
\clearpage

\section*{Methods}

\noindent \textbf{Experiment.} The full experimental details are given in the Supplementary Information. 
\vspace{0.5cm}

\noindent \textbf{Concurrence and Fidelity.} To quantify the quality and degree of entanglement of our we states, we used the Fidelity and Concurrence as our figures of merit.  The fidelity was calculated from
\begin{equation}
    F =\left( \text{Tr}  \left( \sqrt{  \sqrt{\rho_T}\rho_M \sqrt{\rho_T}  }  \right) \right)^2 ,
\end{equation}
where $\rho_T$ is the target density matrix while $\rho_M$ is the measured density matrix. The fidelity is 0 if the states are not identical or 1 when they are identical up to a global phase.  The concurrence was used to measure the degree of entanglement between the hybrid entangled photons. It was calculated from 

\begin{equation}
    C(\rho) = \text{max} \{ 0, \lambda_1 -\lambda_2- \lambda_3 - \lambda_4 \},
\end{equation}

where $\lambda_i$ are eigenvalues of the operator $ R = \text{Tr} \left( \sqrt{  \sqrt{\rho} \tilde{\rho} \sqrt{\rho}  }  \right)$ in descending order and $\tilde{\rho} = \sigma_{y} \otimes \sigma_{y} \rho^* \sigma_{y} \otimes \sigma_{y}$. The concurrence ranges from 0 for separable states to 1 for entangled states.
\vspace{0.5cm}

\noindent \textbf{Quantum Stokes measurements.} The aim is to derive the nonlocal stokes parameters  $\bar{S}( \bar{r} ) = \langle S_x ( \bar{r} ), S_y ( \bar{r} ), S_z ( \bar{r} ) \rangle$. Traditionally, the spatially resolved stokes parameters of a vectorial field are measured from the Pauli matrices in the polarisation degree of freedom, i.e $S_i( \bar{r} )  = \langle \sigma_i \rangle ( \bar{r} )$. 

Here, we achieve this by extracting them from the reconstructed density matrix of the two photon state. We can therefore compute the stokes parameters as
\begin{equation}
    S_j = \text{Tr} \left( \mathbb{I}_{A} \otimes  \sigma_{B,j} \rho \right),
\end{equation}
where $\text{Tr} \left( \cdot \right)$ is the trace operator. Since the general decomposition of the density matrix is given as 
\begin{align}
	\rho & = \sum_{pqst =1} ^{2} \gamma_{pqst} \ket{\ell_p}_A\bra{\ell_q}_A  \otimes  \ket{e_s}_B\bra{e_t}_B ,
	\label{eq:quantumStokes}
\end{align}
where $\gamma_{pqrs}$ are coefficients and $\ket{\ell_{p(q)}}_A$ and $\ket{e_{s(t)}}_B$ are the OAM and polarisation basis states of photon A and B, respectively. It follows that we can now express non-local stokes parameters as
\begin{align}
    S_j = \sum_{pqst =1} ^{2} \gamma_{pqst} \text{Tr} \left( \ket{\ell_p}_A \bra{\ell_q}_A \right)\  \text{Tr} \left( \sigma_{B, j} \ket{e_s}_B\bra{e_t}_B \right).
    \label{eq: expandedStokes}
\end{align}
Next, we apply the trace of photon A in the position basis, $\{ \ket{\bar{r}}_A \ | \ \bar{r} \in \mathcal{R}_A^2 \}$, satisfying the orthogonality ($ \braket{\bar{r}_1 | \bar{r}_2} = \delta \left( \bar{r}_1 - \bar{r}_2 \right)$) and the completeness relation 
($\int  \ket{ \bar{r}}_A \bra{\bar{r}}_A d^2r =\mathbb{I}_A$).
By noting that the OAM eigenmodes can be projected onto the position basis, $\braket{ \bar{r} | \ell } = \text{LG}_{\ell} \left( \bar{r} \right)$, we can perform the trace operation for photon A in Eq. (\ref{eq: expandedStokes}) resulting in
\begin{align}
    S_j(\bar{r}) = \sum_{pqst =1} ^{2} \text{LG}_{\ell_p} \left( \bar{r} \right) \text{LG}^*_{\ell_q} \left( \bar{r} \right) \  \text{Tr}( \sigma_{B, j} \ket{e_r}_B \bra{e_s}_B ).
\end{align}

\vspace{0.5cm}
\noindent \textbf{Quantum Skyrme Number.} To find the Skyrme number, we reconstructed the paraxial skyrmion field, $\Sigma_z$, using the quantum stokes parameters

\begin{equation}
    \Sigma_z (x,y)= \frac{1}{2} \epsilon_{pqr} S_p \frac{\partial S_q}{\partial x} \frac{\partial S_r}{\partial y},
\end{equation}

where $(p,q,r) = (x,y,z)$ and $\epsilon_{ijk}$ is the Levi-Cevita tensor. Since the Stokes parameters expressed in the position basis are spatially dependent on cylindrically symmetric Laguerre Gaussian functions, the Skyrme number was calculated using

\begin{equation}
    N = \frac{1}{4\pi}\int\limits_{0}^{\infty}\int\limits_{0}^{2\pi} \Sigma_z  d\varphi dr \, .
    \label{eqn: SkyNum}
\end{equation}

\section*{Acknowledgements}
This work was supported by the South African National Research Foundation/CSIR Rental Pool Programme.

\section*{Author contributions}
The experiment was performed by P.O. and I.N. performed the experiment, and P.O., I.N. and R.M.K. contributed the theory.  All authors contributed to the writing of the manuscript and analysis of data. A.F. conceived of the idea and supervised the project.

\section*{Competing Interests}
The authors declare no competing interests.

\section*{Data availability}
The data are is available from the corresponding author on request.

\newpage

\newpage
\clearpage

\section{Supplementary: Quantum topological invariance}
Topology is coded into the global features of the wave function so that, intuitively, a perturbation affecting local features of a wavefunction, should not affect said wavefunction's topology. Thereby making information encoded in the topology robust against such perturbations. We can do better than appealing to intuition: there is a simple mechanism which ensures that topologically encoded information is resilient to perturbations. In this section we outline this mechanism.

Topology is a systematic approach to characterizing those quantities that are insensitive to smooth deformation. In our problem, this is the statement that the skyrmion number is unchanged by a change of coordinates $\mathbf{r}_A$ for photon A. To see this, start from the formula for the skyrmion number
\begin{equation}
    N = \frac{1}{4\pi}\int\limits_{-\infty}^{\infty}\!\int\limits_{-\infty}^{\infty} \frac{1}{2}\epsilon_{zqr} S_p \frac{\partial S_q}{\partial x} \frac{\partial S_r}{\partial y}  dx dy.  \label{eqn:TopInvSkyNum}\,
\end{equation}
Consider a change of coordinates from $x,y$ to $x',y'$ where both are arbitrary functions of $x,y$. First consider the transformation of the integrand under this change of coordinates. A simple application of the chain rule gives
\begin{eqnarray}
&&\frac{1}{2} \epsilon_{pqr} S_p \frac{\partial S_q}{\partial x} \frac{\partial S_r}{\partial y}\cr\cr\cr&=& \frac{1}{2} \epsilon_{pqr} S_p \left(\frac{\partial x'}{\partial x}\frac{\partial S_q}{\partial x'}+\frac{\partial y'}{\partial x}\frac{\partial S_q}{\partial y'}\right)
\left(\frac{\partial x'}{\partial y}\frac{\partial S_r}{\partial x'}+\frac{\partial y'}{\partial y}\frac{\partial S_r}{\partial y'}\right)\cr\cr\cr
&=& \frac{1}{2} \epsilon_{pqr} S_p \left(\frac{\partial x'}{\partial x}\frac{\partial x'}{\partial y}\frac{\partial S_q}{\partial x'}\frac{\partial S_r}{\partial x'}+\frac{\partial x'}{\partial x}\frac{\partial y'}{\partial y}\frac{\partial S_q}{\partial x'}\frac{\partial S_r}{\partial y'}\right.\cr\cr
&&\qquad\qquad\qquad\left.+\frac{\partial y'}{\partial x}\frac{\partial x'}{\partial y}\frac{\partial S_q}{\partial y'}\frac{\partial S_r}{\partial x'}+\frac{\partial y'}{\partial x}\frac{\partial y'}{\partial y}\frac{\partial S_q}{\partial y'}\frac{\partial S_r}{\partial y'}\right).\cr\cr&&
\end{eqnarray}
There are four terms in the final expression above. The indices $q$ and $r$ are summed. The first and fourth terms above are sums of an expression that is symmetric under interchange of $q$ and $r$, times $\epsilon_{zqr}$ which is antisymmetric under the interchange of $q$ and $r$. Consequently the first and fourth terms vanish after the sums over $q$ and $r$ is performed. In the third term we relabel the indices $q\leftrightarrow r$ to obtain
\begin{eqnarray}
&&\frac{1}{2} \epsilon_{pqr} S_p \frac{\partial S_q}{\partial x} \frac{\partial S_r}{\partial y}\cr\cr
&=& \left(\frac{\partial x'}{\partial x}\frac{\partial y'}{\partial y}-\frac{\partial y'}{\partial x}\frac{\partial x'}{\partial y}\right) \frac{1}{2} \epsilon_{pqr} S_p \frac{\partial S_q}{\partial x'}\frac{\partial S_r}{\partial y'}.\cr\cr
&&
\end{eqnarray}
The integration measure transforms as usual
\begin{equation}
dx dy =J dx' dy',
\end{equation}
where the Jacobian is given by
\begin{eqnarray}
J&=&\left(\frac{\partial x}{\partial x'}\frac{\partial y}{\partial y'}-\frac{\partial y}{\partial x'}\frac{\partial x}{\partial y'}\right)\cr\cr
&=&\left(\frac{\partial x'}{\partial x}\frac{\partial y'}{\partial y}-\frac{\partial y'}{\partial x}\frac{\partial x'}{\partial y}\right)^{-1}.
\end{eqnarray}
Putting these transformation rules together we learn that
\begin{eqnarray}
    N &=& \frac{1}{4\pi}\int\limits_{-\infty}^{\infty}\!\int\limits_{-\infty}^{\infty} 
\frac{1}{2} \epsilon_{pqr} S_p \frac{\partial S_q}{\partial x} \frac{\partial S_r}{\partial y}  dx dy\cr\cr
      &=& \frac{1}{4\pi}\int\limits_{-\infty}^{\infty}\!\int\limits_{-\infty}^{\infty} \frac{1}{2} \epsilon_{pqr} S_p \frac{\partial S_q}{\partial x'} \frac{\partial S_r}{\partial y'} dx' dy',
\end{eqnarray}
which tells us that the skyrmion number is independent of the choice of coordinates used for the calculation. This is the complete content of the statement that the skyrmion number is invariant under smooth deformations, since any smooth deformation is equivalent to a judiciously chosen change of coordinate. 

\begin{figure}[h]
\includegraphics[width=0.9\linewidth]{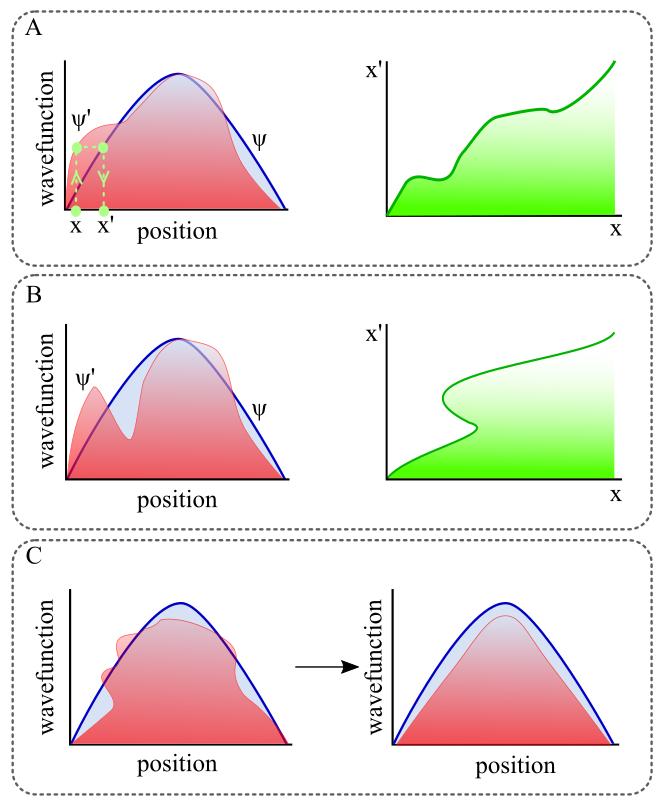}
\caption{\textbf{The origin of topologically protected wavefunctions.} This figure describes the coordinate transformation defined by the original and noisy wavefunctions. To obtain an $x'$ starting from a given $x$, as in (A) we follow point $x$ vertically up to $\psi'$, then traverse horizontally from $\psi'$ to $\psi$ and finally project down from $\psi$ to $x'$. If the noise adds extra local maxima and minima, as shown in (B), the map between $x$ and $x'$ will zig-zag as shown. Finally, the noise can introduce a discrepancy between the amplitude of $\psi$ and the amplitude of $\psi'$. In this case we solve the relation $\psi'(x)=A\psi(x')$, with the constant $A$ magnifying $\psi$ to match $\psi'$. Since wavefunctions are rays in Hilbert space, two wavefunctions that differ only in magnitude and/or global phase, are the same state. Consequently $\psi'(x)=A\psi(x')$ continues to identify the original and noisy states correctly.}
\label{fig:coordtransform}
\end{figure}

To complete the argument, we need to explain why the freedom to change coordinates leads to resilience against local perturbations. The argument is most simply developed in a single spatial dimension. In the presence of a perturbation, the wave function $\psi(x)$ is changed into $\psi'(x)$. If the perturbation is not too severe, the equality $\psi'(x)=\psi(x')$ defines a change of coordinates from $x$ to a new coordinate $x'$. See Figure~\ref{fig:coordtransform}A which shows an example of this coordinate transformation. We are interested in computing a topological invariant associated with the distorted wave function. Since the topological invariant is unchanged by coordinate transformations, we are free to perform the transformation $x\to x'$ that we have just defined. The transformation replaces $\psi'(x)\to\psi(x')$, so that we return to our undistorted wave function in the new coordinates, demonstrating the topological perturbation invariance mechanism. In Figure~\ref{fig:coordtransform}A the perturbation was not severe enough to introduce new maxima and minima in the wave function. In Figure~\ref{fig:coordtransform}B perturbation with a greater amplitude is applied, leading to a new local maximum and a new local minimum. The equation $\psi'(x)=\psi(x')$ continues to define a change of coordinates from $x$ to the new coordinate $x'$, but now the path must backtrack to produce the maxima and minima, so that the transformation between $x$ and $x'$ zig-zags as shown. This is still a perfectly smooth change of coordinate and the topological invariant remains unchanged. Finally, as shown in Figure~\ref{fig:coordtransform}C, we might induce such a large perturbation, i.e severely distort the wave function such that there is no coordinate transformation that manages to completely restore the wave function. In this case the best we can do is to require that $\psi'(x)=A\psi(x')$, where both the relation between $x$ and $x'$ as well as the amplitude $A$ are parameters we can vary. Recall that physical states are rays in Hilbert space. Two wavefunctions that differ only in magnitude and/or global phase, correspond to the same ray and consequently they define the same physical state. Thus the condition $\psi'(x)=A\psi(x')$ continues to identify the original and perturbed states correctly. As the amplitude of the perturbation is increased, we will not manage to satisfy the condition $\psi'(x)=A\psi(x')$ exactly, and the perturbation invariance is no longer perfect. Ultimately, if the perturbation is to severe, there is no coordinate transformation that even approximately obeys $\psi'(x)=\psi(x')$. At this point we can no longer smoothly deform $\psi'$ into $\psi$, i.e. the two wave functions have different topologies and topological perturbation invariance will fail. At these very high perturbation levels even basic properties, like the number of nodes of the wave function, are changed.

Now that the basic idea is clear, we will describe how this argument applies to the actual problem of interest. Recall that the wave function, after expanding in the basis of position space states $\ket{\mathbf{r}_A}$, is given by
\begin{align}
 \ket{\Psi} = \int_{ \mathcal{R}^2 }  \ket{\mathbf{r}_A} \left( a(\mathbf{r}_A) \ket{H}_B + b(\mathbf{r}_A) e^{ i\Theta(\phi_A)}\ket{V}_B \right) d^2 r_A,  
 \label{eq: SuppPosBasis}
 \end{align}
\noindent 
where $a(\mathbf{r}_A) \equiv |\text{LG}_{\ell_{1}} \left( \mathbf{r}_A \right)|$, $b(\mathbf{r}_A) \equiv |\text{LG}_{\ell_{2}} \left( \mathbf{r}_A \right)|$, $\Theta(\phi_A) = \Delta\ell\phi_A + \gamma$ and $\Delta\ell = \ell_2-\ell_1$. As we have remarked, the coefficient of $\ket{\mathbf{r}_A}$ above defines a state for photon $B$
\begin{align}
 \ket{\psi_{B|A}} = \cos(\theta(\mathbf{r}_A)) \ket{H}_B +  \sin(\theta(\mathbf{r}_A)) e^{ i\Theta(\phi_A)}\ket{V}_B \,.
\end{align}
The $\theta(\mathbf{r}_A)$ dependence arises since we have normalized the state. Giving up normalization, we can write this state as
\begin{align}
 \ket{\psi_{B|A}}\propto\ket{H}_B+\tan(\theta(\mathbf{r}_A))e^{i\Theta(\phi_A)}\ket{V}_B .
\end{align}
Both the entanglement structure of this state as well as its topology i.e. its skyrmion number, are completely specified by the coefficient $\tan(\theta(\mathbf{r}_A))e^{i\Theta(\phi_A)}$. In the presence of some perturbation this coefficient is modified to $\tan(\theta'(\mathbf{r}_A))e^{i\Theta'(\phi_A)}$. The coordinate transformation which establishes that both the entanglement of the wave function and it's topology are robust against against local perturbations is defined by
\begin{equation}
\tan(\theta'(\mathbf{r}_A))e^{i\Theta'(\phi_A)}=\tan(\theta(\mathbf{r}'_A))e^{i\Theta(\phi_A')}.
\label{eq: TopRobust}
\end{equation}
This coordinate transformation protects the relative weighting of the two components of the wave function, ensuring that it does indeed protect both the entanglement and topology of the wave function.

\subsection{Topological resilience against entanglement decay}

\begin{figure}[h]
\includegraphics[width=1\linewidth]{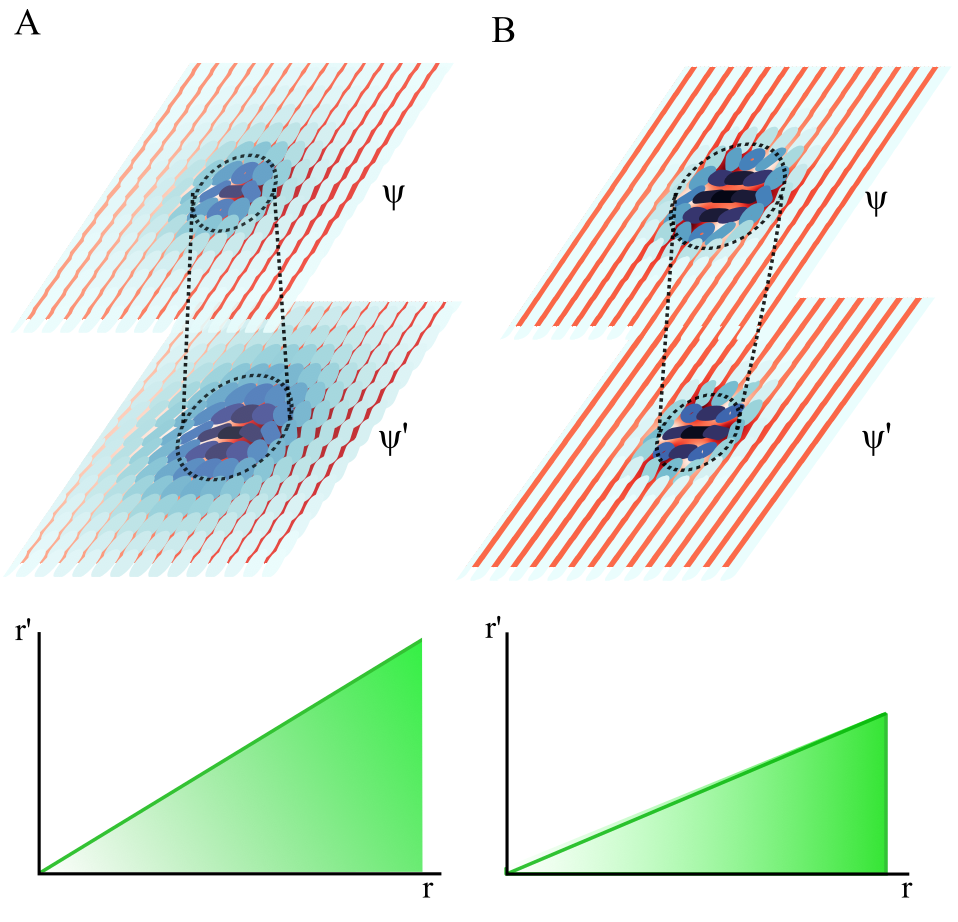}
\caption{\textbf{Entanglement Decay as smooth deformation} The entanglement decay operation, $\hat{T}_d$, can be described as a smooth deformation as it is does not affect the topology of the quantum state. (A) Physically, if we consider any biphoton state with $|N| = 1$, the action of the operation on the state works to shift all polarizations states towards or away from the centre such that one can perform a simple coordinate change on the radial position coordinate, i.e $\ket{\Psi(r_A,\phi_A)} \xrightarrow{\text{$\hat{T}_d$}} \ket{\Psi'(r_A,\phi_A)} = \ket{\Psi(r'_A,\phi_A)}$. (B) For any biphoton state with $|N| > 1$, the same coordinate transformation discussed above is valid, except that the coordinate tranformation follows a slightly different rule which depends on $|\ell_2| - |\ell_1|$.}
\label{fig:EntDecaycoordtransform}
\end{figure}

We consider the experimentally demonstrated example of the topology preserving entanglement decay operation, $\hat{T}_d$, and describe it topologically as a simple coordinate transformation as it preserves the topology of a given wavefunction so long as the new wavefunction describes an entangled state. Suppose we start with the entangled state, $\ket{\Psi (a(r_A), b(r_A), \Theta(\phi_A))}$ given in equation \ref{eq: SuppPosBasis} with a possible example topologies given in Figure~\ref{fig:EntDecaycoordtransform}A and B, and we apply an entanglement decay operation on this state such that it is changed to $\ket{\Psi' (a'(r_A), b'(r_A), \Theta(\phi_A))}$ with $a'(r_A) = \alpha a(r_A)$, $b'(r_A) = \sqrt{1 - \alpha^2}b(r_A)$ and $\alpha \in (0,1)$.

To equate this transformation to a coordinate transformation, we consider what happens to its structure locally. In other words, suppose that under the state $\ket{\Psi}$, a spatial measurement on photon A, at $\mathbf{r_A} = (r_A,\phi_A)$ results in a collapse of photon B into the polarization state $\ket{P}$, and under the state $\ket{\Psi'}$, a spatial measurement on photon A, at $\mathbf{r'_A} = (r'_A,\phi'_A)$ results in a collapse of photon B into the polarization state $\ket{P}$. If a coordinate mapping, $\mathbf{r_A} \to \mathbf{r'_A}$ exists such that the scenario above is satisfied for all $\mathbf{r_A},\mathbf{r'_A} \in \mathbb{R}$, then we can define a smooth deformation describing the action of $\hat{T}_d$ on our wavefunction.

Lending from the discussion above and using a modified version of equation \ref{eq: TopRobust}, we have that a coordinate transformation must exist such that

\begin{align}
    \frac{b(r_A)}{a(r_A)} \xrightarrow{\text{$\hat{T}_d$}} \frac{b'(r_A)}{a'(r_A)} = \frac{b(r'_A)}{a(r'_A)}
    \label{eq: TopCond}
\end{align}

where the terms dependent on the azimuthal coordinate $\phi_A$ were factored out of the expression as the transformation did not modify them. Substituting the appropriate Laguerre Gaussian functions into equation \ref{eq: TopCond} yields

\begin{eqnarray}
    \frac{b'(r_A)}{a'(r_A)} &=& \sqrt{\frac{|\ell_2|!}{|\ell_1|!}} \left[\frac{\sqrt{2}}{\omega_0} r_A \left(\frac{\sqrt{1-\alpha^2}}{\alpha}\right)^{\frac{1}{|\ell_2|-|\ell_1|}}\right]^{|\ell_2|-|\ell_1|} \cr\cr\cr &=&\sqrt{\frac{|\ell_2|!}{|\ell_1|!}} \left[\frac{\sqrt{2}}{\omega_0} r'_A \right]^{|\ell_2|-|\ell_1|} = \frac{b(r'_A)}{a(r'_A)},
    \label{eq: TopCond}
\end{eqnarray}

from which we extract the coordinate transformation $(r_A, \phi_A) \to (r'_A, \phi'_A)$ where $r'_A = r_A \left(\frac{\sqrt{1-\alpha^2}}{\alpha}\right)^{\frac{1}{|\ell_2|-|\ell_1|}}$ and $\phi'_A = \phi_A$.

\section{Supplementary: Experiment}
The creation of skyrmionic entangled states can be broken up into three parts, the Generation, Spatial-to-Polarization Conversion (SPC) and Detection. The experimental schematic is shown in figure \ref{fig:ExpFig} A. We start by generating an entangled pair of photons through Spontaneous Parametric Down-Conversion (SPDC), by sending a magnified ($f_1 = 50$ mm and $f_2 = 25$ mm) 405nm wavelength laser beam through a Non-linear PPKTP (Type I) Crystal (NC). By using a Band-Pass Filter (BPF) we are left only with the down-converted 810nm wavelength light. Through this process, we generate photon pairs entangled according to the non-separable wavefunction

\begin{equation}
    | \Psi \rangle = \sum_{\ell} c_{\ell} \ket{\ell}_A \otimes \ket{-\ell}_B. 
    \label{eq: suppOAMEnt}
\end{equation}

Magnfication of the beam incident on the NC is required to flatten the spiral bandwidth of the NC thereby allowing access to higher order Orbital Angular Momentum (OAM) modes \cite{miatto2012spatial, Miatto2012}. The down-converted light is then rotated to horizontal polarization for modulation by a Spatial Light Modulator (SLM). The entangled pair of photons, photons A and B, are spatially separated using a 50:50 Beam-Splitter (BS). Photon A, which will ultimately carry the spatial DoF, is detected through a combination of an SLM and a Single-Mode Fibre (SMF) coupled to an Avalanche Photon Detector (APD). To achieve our topologically non-trivial states, we make use of hybrid entangled photons, which means that these photons are entangled in different Degrees of Freedom (DoF). In our case, while photon A carries the spatial/OAM DoF, we couple photon B's Spatial information to polarization such that the modified wave function of the entangled pair is of the form

\begin{equation}
    | \Psi \rangle = \alpha|\ell_1\rangle_A |H\rangle_B + \sqrt{1-\alpha^2}e^{i\gamma}|\ell_2\rangle_A |V\rangle_B, 
    \label{eq: suppHybEnt}
\end{equation}

where $\ket{H}, \ket{V}$ are the usual orthogonal horizontal and vertical polarization states, respectively, $\alpha \in [0,1]$ is a weighting parameter controlling the degree of entanglement between photons A and B and $\gamma$ is a relative phase between the terms of the wavefunction. To engineer a desired state with a non-trivial entanglement topology, we require full control not only of the parameters $\alpha$ and $\gamma$ but also of the desired OAM subspace. For this, typical methods used to generate hybrid entangled states using q-plates \cite{Marrucci_2011} and static meta-surfaces \cite{Balthasar_2017, Robert_2017} will not suffice, hence we make use of an all-digital SPC approach \cite{Nape_2022} which couples the OAM of photon B to a particular polarization state. To do this we send photon B through a Half-Wave Plate (HWP) to convert it to diagonal polarization before having its OAM information coupled to polarization, thereby transferring the OAM-OAM correlations between photons A and B to OAM-Polarization correlations. Furthermore, holograms of asymmetric azimuthal charge are displayed on the SLM, in order to post-select the desired asymmetric entangled state which yields non-trivial topologies. The SPC process occurs through a double bounce where on the first bounce, horizontally polarized photons are modulated and after passing through the Quarter-Wave Plate (QWP) and mirror configuration shown in the schematic, we have that on the second bounce, the left-over light in the orthogonal polarization gets modulated. Mathematically, this SPC conversion can be written as a mapping of the form \cite{Nape_2022}

\begin{eqnarray}
        h_{SPC} &:& \ket{\ell'}_B\ket{V}_B \to \alpha \ket{0}_B \ket{H}_B\\
        &:& \ket{\ell ''}_B\ket{H}_B \to \sqrt{1-\alpha^2} e^{i\gamma}\ket{0}_B\ket{V}_B,
\end{eqnarray}

where $\ell'$ and $\ell''$ are any two post-selected OAM states. Photon B is then measured using a set of polarization optics, a HWP orientated to
$45^{\circ}$ and a Linear Polarizer orientated at $90^{\circ}$. Photons A and B are measured in coincidence allowing for a full Quantum State Tomography to be performed in order to reconstruct the state. Figures \ref{fig:ExpFig} B and C show the different holograms that are displayed to capture all the data for a single quantum state, namely the generated state $\ket{\Psi} = \ket{3}_A\ket{H}_B + \ket{0}_A\ket{V}_B$. SLM A cycles through 6 holograms, projecting photon A onto the spatial modes associated with $\ket{3}, \ket{0}$ as well as their superpositions given by $\ket{\theta} = \{\frac{1}{\sqrt{2}} \left( \ket{0}_A\pm \exp(i\theta_{A})\ket{3}_A \right) \}, \theta =  (0, \pi/2)$. SLM B cycles through 6 holograms, projecting photon B onto the polarization modes associated with $\{\ket{H}, \ket{V}\}$ as well as their superpositions which are given by the polarization states $\{\ket{D}, \ket{A}, \ket{R}, \ket{L} \}$. $\gamma$ can be controlled rather simply by changing the global phase on one of the SLM halves such that during the SPC process, only one term in the wavefunction sees this phase. Finally, in order to control $\alpha$ we introduce a loss in the SPC conversion. Since the double-bounce mechanism necessitates that the horizontally and vertically polarized photons follow the same path, we can induce a loss in the system by sending a precise portion of light away from the experimental path. This is done by adding a grating onto one of the halves of the SLM (see figure \ref{fig:ExpFig} C) and controlling the grating depth. In modifying the grating depth, we modify the efficiency of the SLM in the nth order, which is given by \cite{Rosales2017SLM}

\begin{equation}
    |c_n|^2 = \mathrm{sinc}^2(\pi(n-M)),
\end{equation}

where $|c_n|^2$ is the fraction of power in the n-th diffraction order and $M\in[0,1]$ controls the grating depth from $0$ to $2M\pi$. Since the SPC process used keeps the orthogonal polarizations collinear, $\alpha$ is dependent on the fraction of power sent to the zeroth order. Since the induced loss was to the vertically polarized term in equation \ref{eq: suppHybEnt}, we have that $\alpha = 0.5 (1)$ when $M = 0 (1)$. As such we can show that $\alpha$ is dependent on $M$ according to the expression

\begin{equation}
    \alpha(M) \propto \frac{\sqrt{1 - |c_0|^2}}{2} + \frac{1}{2}.
\end{equation}

This establishes a relation between $\alpha$ and the grating depth, more specifically by varying the grating depth, we are able to vary the degree of entanglement, with maximal entanglement obtained when the grating is switched off, and no entanglement when the grating is switched on with a maximum depth. Experimentally however, we vary the grating depth non-linearly between a grating depth of $0(M=0)$ and $2\pi(M=1)$ and extract $\alpha$ through Concurrence and Fidelity calculations.\\

\begin{figure*}[t!]
\includegraphics[width=\textwidth]{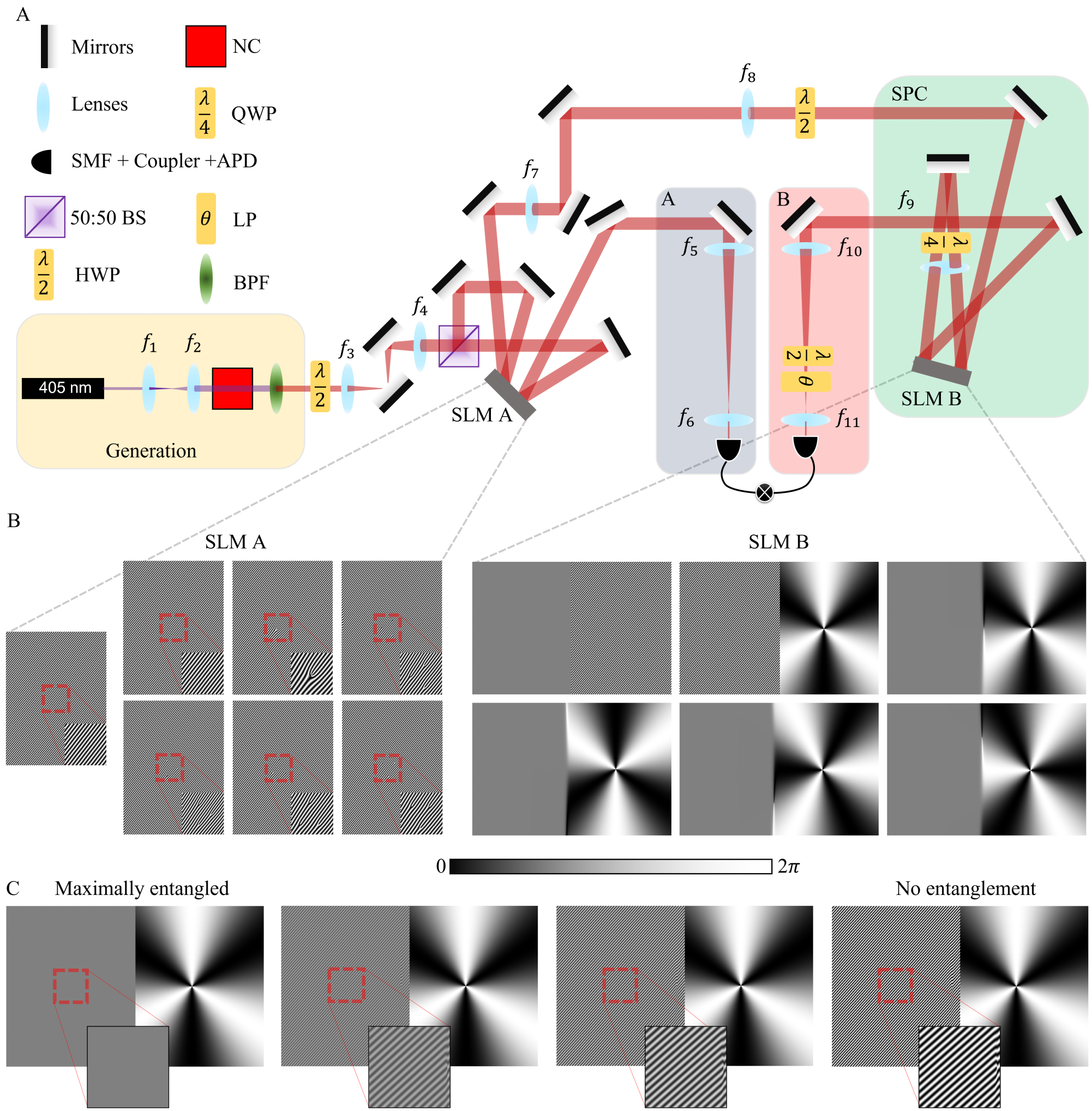}
\caption{(A) Experimental schematic for generation, and detection of topological hybrid entangled states.  An incident 405nm wavelength pump beam produced from an OBIS 405LX diode laser is incident on a temperature controlled type 1 PPKTP Non-linear Crystal (NC) producing pairs of entangled photons of wavelength 810nm. Any residual pump signal is then filtered out using a Band-pass filter (BPF) and the remaining signal and idler photons are sent through a Half-Wave Plate (HWP) orientated to ensure the photons are horizontally polarized. Entangled photons are then spatially separated via a 50:50 beamsplitter. Photon A is measured using a HOLOEYE PLUTO-2.1 Spatial Light Modulator (SLM A), Single Mode Fiber (SMF) and Perkin Elmer - SPCM-13 Avalanche Photon Detector (APD). Photon B is sent through a digital Spatial-to-Polarization Converter (SPC). Photon B is then measured using polarization optics, Half-Wave Plate (HWP) and Linear Polarizer (LP), SMF and APD. The entangled photons are measured in coincidence with the SMFs coupled to single-photon detectors. (B) The holograms displayed on SLM A and SLM B to perform projective measurements on photon A and B, respectively. SLMs A and B are both divided into two halves, where the LHS hologram of SLM A is given by the single panel on the far left, whereas the 6 holograms which the RHS of SLM will cycle through is given by the 6 panels to the right of the LHS panel. (C) To control the degree of entanglement, a grating is placed on one of the halves of the SLM in the SPC conversion with the grating depth controlling the level of entanglement, from no grating leading to maximal entanglement and maximum grating depth leading to no entanglement.}
\label{fig:ExpFig}
\end{figure*}

\section{Supplementary: Quantum State Tomography}

\begin{figure}[t]
\includegraphics[width=\linewidth]{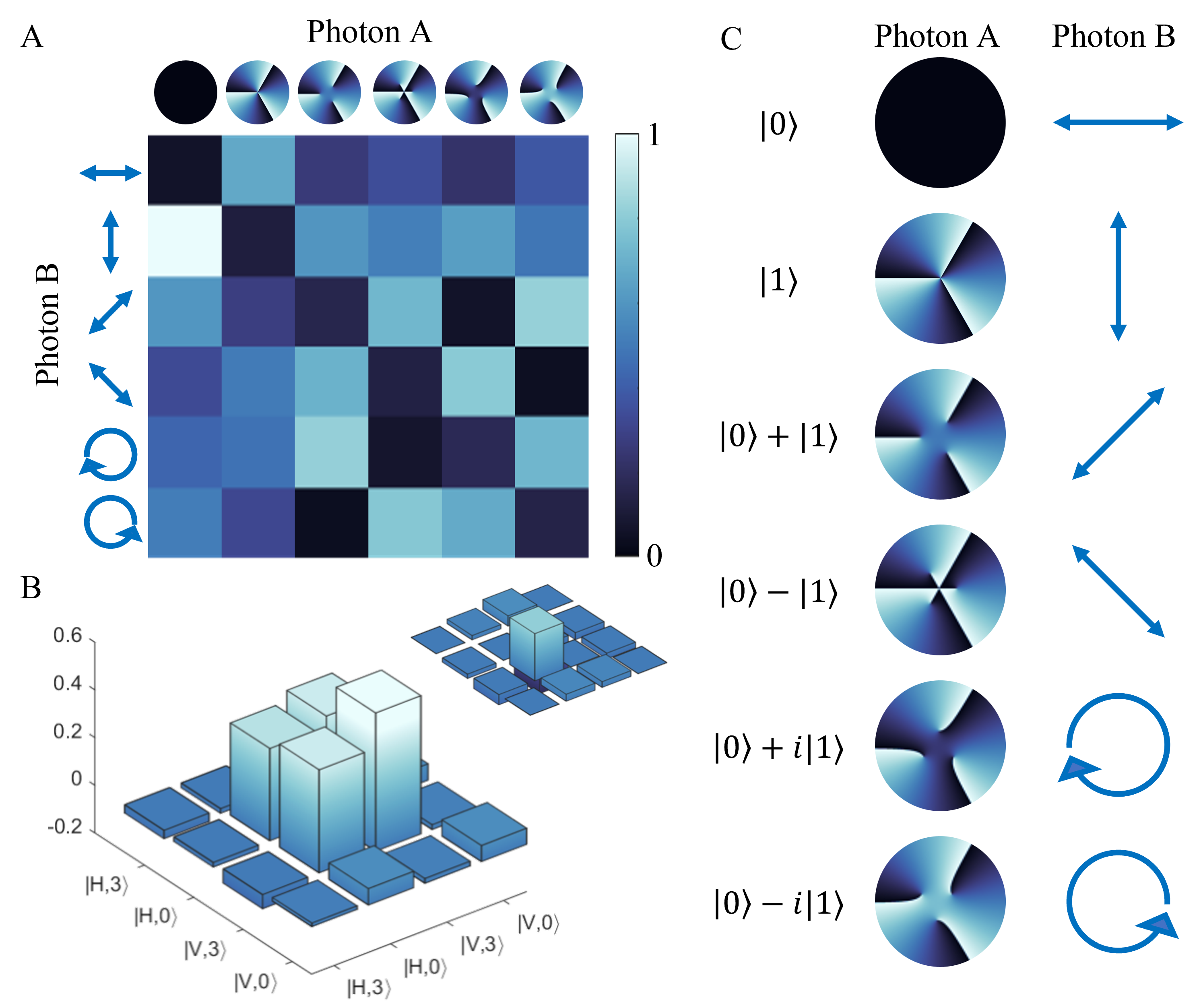}
\caption{(A) Experimental example of QST and (B) reconstructed density matrix for the experimentally generated state, $\ket{\Psi} = \ket{3}_A\ket{H} + \ket{0}_A\ket{V}$. (A) Measured probabilities are shown for spatial (across each row) and polarization (down each column) projections on photon A and B, respectively. Top left 2x2 block shows approximate state where the probability Photon A is in the state $\ket{3}$ given that Photon B is measured to be in the state $\ket{H} (\ket{V})$ is almost $100 \% (0 \%)$, whereas the probability of photon A being in the state $\ket{0}$ is $0\% (100 \%)$, which indicates that $\Psi$ is indeed at least an approximation of our final state. The remaining information provided by the QST gives insight into the purity of the generated state. This tomography is then used to generate (B) the real and imaginary parts (inset) of the density matrix which can then be mapped uniquely (up to some global phase) to the generated biphoton state. From the generated density matrix, one can see some small non-zero contributions appearing outside the central 2x2 matrix which contribute to impurity in the measured state, however this particular density matrix was identified to have a high fidelity of $91.61 \%$ when compared to the density matrix of the intended, ideal pure state. (C) The projective measurement states for photon A and B written in terms of the computational basis states and superpositions of said states.}
\label{fig:QSTEx}
\end{figure}

We reconstructed our two photon nonlocal Skyrmionic field by performing Quantum State Tomography in the modal space in order to obtain the density matrix, $\rho$, of the system. An example of a generated QST and associated density matrix is shown in figure \ref{fig:QSTEx}, derived from data captured for the state $\ket{\Psi} = \ket{3}_A \ket{H}_B + \ket{0}_A \ket{V}_B$. To achieve this, we performed spatially separated projective measurements, $M_{ij}=P^{A}_{i} \otimes P^{B}_{j}$, on each particle, where $P^{A}_{i}$ and $P^{B}_{j}$ are local projections of photon A and photon B in their independent spatial and polarisation DoF, respectively. 


In the computational basis, $\{ \ket{0}, \ket{1} \}$, each of the photons are projected onto the eigenstates $\ket{0}$ and $\ket{1}$ as well as superposition states that form the mutually unbiased basis, $\{ \frac{1}{\sqrt{2}} \left( \ket{0}\pm \exp(i\theta_{A(B)})\ket{1} \right) \}, \theta =  (0, \pi/2)$, with respect to the standard basis. For photon A, this corresponds to the OAM eigenstates, $\ket{\ell_{1, 2}}_A$ and the mutually unbiased superpositions $\ket{\theta} = \frac{1}{\sqrt{2}} \left( \ket{\ell_1}_A + \exp \left( i \theta_A \right) \ket{\ell_2}_A \right)$  which constitute our spatial basis. 

As for photon B, we have the linear horizontal ($\ket{H}_B$) and vertical ($\ket{V}_B$) polarisation states as our standard basis and the superposition states comprised of the diagonal ($\ket{D}_B$), anti-diagonal ($\ket{A}_B$), right circular ($\ket{R}_B$) and left circular ($\ket{L}_B$) polarisation states. We performed these projections digitally using SLMs and linear optical elements only. 
In figure \ref{fig:QSTEx} A, the detection probabilities for each possible projective measurement are shown with varying eigenstate superpositions for photon A and B shown across the rows and down the columns, respectively. The entire set of data is gathered digitally by varying the displayed holograms (see figure \ref{fig:QSTEx}) on the SLM and detecting the number of counts measured in coincidence for each configuration. The detection probabilities were used to determine the density matrix (see figure \ref{fig:ExpFig} B) via least squares fitting. In our model, we assumed that the density matrix follows the decomposition
\begin{equation}
	\rho = \frac{1}{4} \big( \mathbb{I}_4 + \sum^{3}_{m,n=1} b_{mn} \sigma_{A,m} \otimes \sigma_{B,n} \big),
\end{equation}
\noindent where $\mathbb{I}_4$ is the four dimensional identity matrix and $b_{mn}$ is the $\sigma_{A,m}$ and $\sigma_{B,n}$ are the Pauli matrices that span the two-dimensional hybrid space for the spatial and polarisation DoF of photon A and B, respectively. Here, the least squares fitting was used to minimise the relative errors between the measured and detection probabilities, given by 
\begin{equation}
    \chi = \sum_{mn} |p_{mn}(\mathbf{b})-p_{mn}^{M}|^2, 
    \label{eq: detProb}
\end{equation}
 where $p_{mn}(\mathbf{b})$ are the detection probabilities for a given density matrix that is determined by the coefficients $b_{mn}$ which are elements of the matrix $\mathbf{b}$ while $p_{mn}^{M}$ are the probabilities that were measured in the experiment.

Upon obtaining the density matrix $\rho$ we can express it in our OAM and polarisation basis 
\begin{align}
	\rho & = \sum_{pqst} \gamma_{pqst} \ket{\ell_p}_A\bra{\ell_q}_A  \otimes  \ket{e_s}_B\bra{e_t}_B,
	\label{eq:quantumDensityMatrix}
\end{align}
where $\gamma_{pqst}$ are the coefficients that determine the state. 

\section{Supplementary: Concurrence and Fidelity}
\noindent To quantify the quality and degree of entanglement of our we states, we used the Fidelity and Concurrence as our figures of merit. 

The fidelity was measured from
\begin{equation}
    F =\left( \text{Tr}  \left( \sqrt{  \sqrt{\rho_T}\rho_M \sqrt{\rho_T}  }  \right) \right)^2 ,
\end{equation}
where $\rho_T$ is the target density matrix while $\rho_M$ is the measured density matrix. The fidelity is 0 if the states are not identical or 1 when they are identical up to a global phase.

The concurrence was used to measure the degree of entanglement between the hybrid entangled photons. It was measured from 

\begin{equation}
    C(\rho) = \text{max} \{ 0, \lambda_1 -\lambda_2- \lambda_3 - \lambda_4 \},
\end{equation}

where $\lambda_i$ are eigenvalues of the operator $ R = \text{Tr} \left( \sqrt{  \sqrt{\rho} \tilde{\rho} \sqrt{\rho}  }  \right)$ in descending order and $\tilde{\rho} = \sigma_{y} \otimes \sigma_{y} \rho^* \sigma_{y} \otimes \sigma_{y}$. The concurrence ranges from 0 for separable states to 1 for entangled states.

\section{Supplementary: Quantum Stokes measurements}

The aim is to derive the nonlocal stokes parameters  $\bar{S}( \bar{r} ) = \langle S_x ( \bar{r} ), S_y ( \bar{r} ), S_z ( \bar{r} ) \rangle$. Traditionally, the spatially resolved stokes parameters of a vectorial field are measured from the Pauli matrices in the polarisation degree of freedom, i.e $S_i( \bar{r} )  = \langle \sigma_i \rangle ( \bar{r} )$. 

Here, we achieve this by extracting them from the reconstructed density matrix of the two photon state. We can therefore compute the stokes parameters as
\begin{equation}
    S_j = \text{Tr} \left( \mathbb{I}_{A} \otimes  \sigma_{B,j} \rho \right),
\end{equation}
where $\text{Tr} \left( \cdot \right)$ is the trace operator. Since the general decomposition of the density matrix is given as 
\begin{align}
	\rho & = \sum_{pqst =1} ^{2} \gamma_{pqst} \ket{\ell_p}_A\bra{\ell_q}_A  \otimes  \ket{e_s}_B\bra{e_t}_B ,
	\label{eq:quantumStokes}
\end{align}
where $\gamma_{pqrs}$ are coefficients and $\ket{\ell_{p(q)}}_A$ and $\ket{e_{s(t)}}_B$ are the OAM and polarisation basis states of photon A and B, respectively. It follows that we can now express non-local stokes parameters as
\begin{align}
    S_j = \sum_{pqst =1} ^{2} \gamma_{pqst} \text{Tr} \left( \ket{\ell_p}_A \bra{\ell_q}_A \right)\  \text{Tr} \left( \sigma_{B, j} \ket{e_s}_B\bra{e_t}_B \right).
    \label{eq: expandedStokes}
\end{align}
Next, we apply the trace of photon A in the position basis, $\{ \ket{\bar{r}}_A \ | \ \bar{r} \in \mathcal{R}_A^2 \}$, satisfying the orthogonality ($ \braket{\bar{r}_1 | \bar{r}_2} = \delta \left( \bar{r}_1 - \bar{r}_2 \right)$) and the completeness relation 
($\int  \ket{ \bar{r}}_A \bra{\bar{r}}_A d^2r =\mathbb{I}_A$).
By noting that the OAM eigenmodes can be projected onto the position basis, $\braket{ \bar{r} | \ell } = \text{LG}_{\ell} \left( \bar{r} \right)$, we can perform the trace operation for photon A in Eq. (\ref{eq: expandedStokes}) resulting in
\begin{align}
    S_j(\bar{r}) = \sum_{pqst =1} ^{2} \text{LG}_{\ell_p} \left( \bar{r} \right) \text{LG}^*_{\ell_q} \left( \bar{r} \right) \  \text{Tr}( \sigma_{B, j} \ket{e_r}_B \bra{e_s}_B ).
\end{align}
Furthermore the factor $\text{Tr}( \sigma_{B, j} \ket{e_s}_A\bra{e_t}_A )$ can be written as two terms following the spectral decomposition of the Pauli matrices, i.e. $\sigma_{B, j} = \lambda_{j}^+ P_j^+ - \lambda_{j}^-P_j^-$ where $P_j ^ \pm = \ket{\lambda^{\pm}_j}\bra{\lambda^{\pm}_j}$ for positive and negative eigenvalues $\lambda^{\pm} =\pm 1$, therefore
\begin{align}
    S_j(\bar{r}) &= \lambda_{j}^+ \sum_{pqst=1}^2 \gamma_{pqst} \text{LG}_{\ell_p} \left( \bar{r} \right) \text{LG}^*_{\ell_q} \left( \bar{r} \right) \  \bra{e_s}_B P_j^+\ket{e_t}_B  \nonumber \\ 
      & \quad  + \lambda_{j}^- \sum_{pqst=1} ^2 \gamma_{pqst} \text{LG}_{\ell_p} \left( \bar{r} \right) \text{LG}^*_{\ell_q} \left( \bar{r} \right) \  \bra{e_s}_B P_j^-\ket{e_t}_B.
\end{align}
The expression can be further simplified by contracting the indices $s$ and $t$, that iterate over the polarisation eigenstates. 
\begin{align}
     S_j(\bar{r}) &= \lambda_{j}^+ \sum_{pq} \tilde{\gamma}^{j+}_{pq} \text{LG}_{\ell_p} \left( \bar{r} \right) \text{LG}^*_{\ell_q} \left( \bar{r} \right) \nonumber \\ &+ \lambda_{j}^- \sum_{pq} \tilde{\gamma}^{j-}_{pq} \text{LG}_{\ell_p} \left( \bar{r} \right) \text{LG}^*_{\ell_q} \left( \bar{r} \right), \nonumber \\
      &= \lambda_{j}^{+} I^+_{j}(\bar{r}) + \lambda_{j}^- I^-_{j}(\bar{r}),
\end{align}
Here the new coefficient, $\tilde{\gamma}^{j+}_{pq}$, is obtained from contracting the indices, $st$, since the overlap probability $\bra{e_s}_B P_j^+\ket{e_t}_B$ is a finite number.\\

\noindent We note here that we performed a unitary rotation on the Stokes parameters in order to re-orientate the vector plots. The unitary rotation takes the form of the operator for a QWP, fast axis orientated at $45^{\circ}$, given by
\begin{equation}
    R_{QWP} = \begin{pmatrix} 1 & i \\ i & 1
\end{pmatrix}.
\end{equation}
It can be shown that such a unitary operation has no effect on the density matrix and furthermore the rotation can be described as a change of coordinates $(x,y,z) \to (x',y',z')$ which topologically speaking is a smooth deformation which we have shown preserves the skyrme number implying that the topology remains unchanged.

\section{Supplementary: Quantum Skyrme Number}


To find the Skyrme number, we reconstructed the paraxial skyrmion field, $\Sigma_z$, using the quantum stokes parameters

\begin{equation}
    \Sigma_z (x,y)= \frac{1}{2} \epsilon_{pqr} S_p \frac{\partial S_q}{\partial x} \frac{\partial S_r}{\partial y},
\end{equation}

where $(p,q,r) = (x,y,z)$ and $\epsilon_{ijk}$ is the Levi-Cevita tensor. Since the Stokes parameters expressed in the position basis are spatially dependent on cylindrically symmetric Laguerre Gaussian functions, the Skyrme number was calculated using

\begin{equation}
    N = \frac{1}{4\pi}\int\limits_{0}^{\infty}\int\limits_{0}^{2\pi} \Sigma_z  d\varphi dr, 
    \label{eqn: SkyNum}\,
\end{equation}

which after substitution of the appropriate LG functions can be shown to simplify down to \cite{gao2020paraxial}

\begin{equation}
    N = \Delta \ell \left(\frac{1}{1 + |g(0)|^2} - \frac{1}{1 + |g(\infty)|^2}\right),
\end{equation}

where $g(\mathbf{r}) = \left(\sqrt{\frac{1-\alpha^2}{\alpha^2}}) \right)\frac{LG_{\ell_2} (\mathbf{r})}{LG_{\ell_1}(\mathbf{r})}e^{i\gamma}$ and $\Delta \ell = \ell_2 - \ell_1$. From this expression, one can see that $N \in \mathbb{Z}$ as expected and $N = m\Delta \ell$ with 
\[ 
m = 
\begin{cases}
    0,& |\ell_1| = |\ell_2|\\
    1,& |\ell_1| > |\ell_2|\\
    -1,& |\ell_1| < |\ell_2|.
\end{cases}
\]

\section{Supplementary: Stereographic Projection}

\begin{figure*}[t!]
\includegraphics[width=\textwidth]{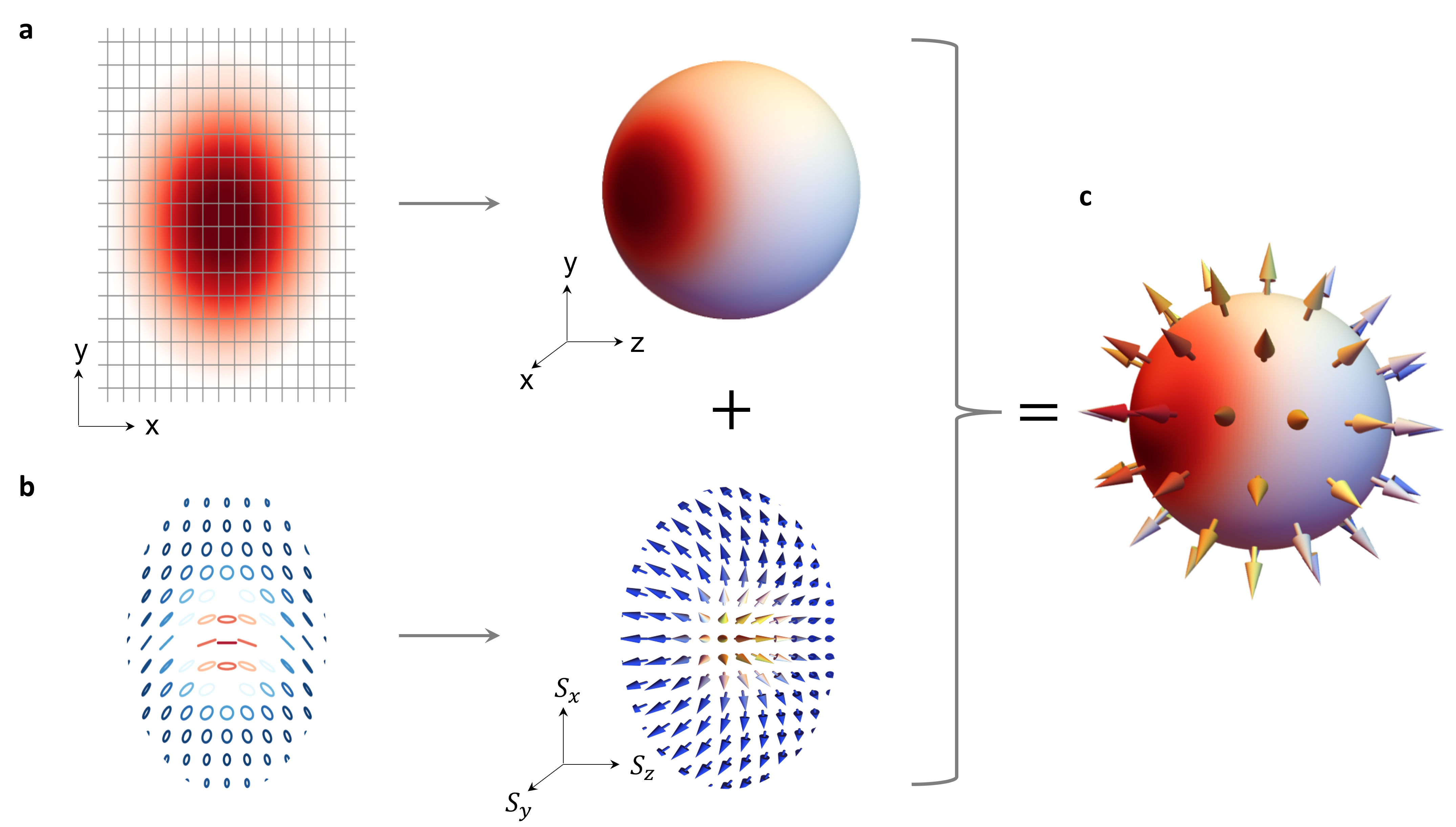}
\caption{\textbf{Stereographic projection of entangled state onto state space} \textbf{a,} Through a stereographic projection, the transverse plane, $\mathcal{R}^2$, is mapped onto the surface of a sphere $\mathcal{S}^2$. \textbf{b,} Arbitrary elliptical polarization states are represented as state vectors pointing from the centre of the Poincar\'e sphere to its surface. \textbf{c,} Combining the stereographic projection of $\mathcal{R}^2$ to $\mathcal{S}^2$ with the exchange of polarization ellipses for Poincar\'e state vectors, we get the condensed representation of the topology of our entangled state.}
\label{fig:Stereo}
\end{figure*}

To compactly represent the topology of an entangled state we may stereographically project the polarization and spatial information onto a single sphere. The spatial projection of $\mathcal{R}^2$ to $\mathcal{S}^2$ is done via the mapping

\begin{equation}
    (x',y',z') = \left( \frac{2x}{1+x^2+y^2}, \frac{2y}{1+x^2+y^2}, \frac{-1+x^2+y^2}{1+x^2+y^2} \right)
\end{equation}

where $(x,y) \in \mathcal{R}^2$ and $(x',y',z') \in \mathcal{S}^2$ with the usual relation $x'^2 + y'^2 + z'^2 = 1$ holding for all $x',y',z'$. An example of this mapping is shown in Figure~\ref{fig:Stereo}\textbf{a}, with additional probability density information also being carried over to the surface of the sphere.  Expressed in polar coordinates the mapping can be written as
\begin{equation}
    (\phi, \theta) = \left( 2 \arctan{\left( \frac{1}{r} \right)}, \Theta \right)
\end{equation}

where $r, \Theta \in \mathcal{R}^2$ and $\phi, \theta \in \mathcal{S}^2$ with $ $. Beyond the inherent compactness of this representation, one is free to scale $r$ and $\Theta$ by arbitrary real constants $a,b^2>0$ which allow for a simple scheme to restructure the representation so as to emphasize regions of interest. The validity of such a restructuring of the representation follows from the fact that the topology is invariant to smooth deformations caused by coordinate changes, and in fact, any scaling can be reversed through a second inverse scaling.

Next, for a clearer depiction of the vectorially textured field formed in the entanglement, we represent the arbitrary elliptical states found in coincidence with spatial measurements on the partner photon, with poincar\'e vector states, as shown in Figure~\ref{fig:Stereo}\textbf{b}. Extracting the Quantum Stokes Parameters from the entangled state, allows for the description of a spatially dependent vector $ \mathbf{S} = \left( S_x (r, \Theta), S_y (r, \Theta), S_z (r, \Theta) \right) $ which carries the local normalization condition $\mathbf{S} \cdot \mathbf{S} = 1$, thus describing a vector field with vectors pointing from the centre of the Poincar\'e sphere to its surface and therefore denoting particular polarization states.

Combining these two representation schemes, we can compactly represent the entirety of the topological information on the surface of a sphere, an example of which is shown in Figure~\ref{fig:Stereo}\textbf{c}. \\ 

\section{Supplementary: Experimental Data}
\begin{table}[h!]
  \centering
    \label{tab:table1}
    \begin{tabular}{c|c|c|c}
      \textbf{Skyrmion} & \textbf{$N_\text{exp}$} & \textbf{$F$} & \textbf{$C$}\\
      \hline
      $\{-3, \frac{\pi}{2}\}$ & -2.999 $\pm 3.292 \times 10^{-5}$ & 0.9724 & 0.8472\\
      $\{-3, 0\}$ & -2.999 $\pm 4.654 \times 10^{-5}$ & 0.9683 & 0.8655\\
      $\{-1, \frac{\pi}{2}\}$ & -0.981 $\pm 5.458 \times 10^{-4}$& 0.9637 & 0.7003\\
      $\{-1, 0 \}$ & -0.982 $\pm 5.087 \times 10^{-4}$& 0.9818 & 0.7351\\
      $\{0, \frac{\pi}{2}\}$ & -0.004 $\pm 1.231 \times 10^{-2}$ & 0.9926 & 0.9147\\
      $\{0, 0\}$ & -0.008 $\pm 8.315 \times 10^{-3}$& 0.9955 & 0.8921\\
      $\{1, \frac{\pi}{2}\}$ & 0.978 $\pm 7.518 \times 10^{-4}$ & 0.9559
      & 0.8438\\
      $\{1, 0\}$ & 0.973 $\pm 5.993 \times 10^{-4}$ & 0.9498 & 0.7381\\
      $\{3, \frac{\pi}{2}\}$ & 2.998 $\pm 3.368 \times 10^{-5}$ & 0.9189 & 0.8346\\
      $\{3, 0\}$ & 2.998 $\pm 4.558 \times 10^{-5}$ & 0.9161 & 0.8165\\
    \end{tabular}
  \caption{Table of experimental data. The Skyrmion topology, $N$, and type $\phi$ is given in the form $\{ N, \phi\}$ along with the associated measured values for the experimental skyrme number, $N_{exp}$, the fidelities of each reconstructed density matrix, $F$, and their concurrences.}
  \label{tbl: Data}
\end{table}

\begin{figure*}[t!]
\includegraphics[width=\textwidth]{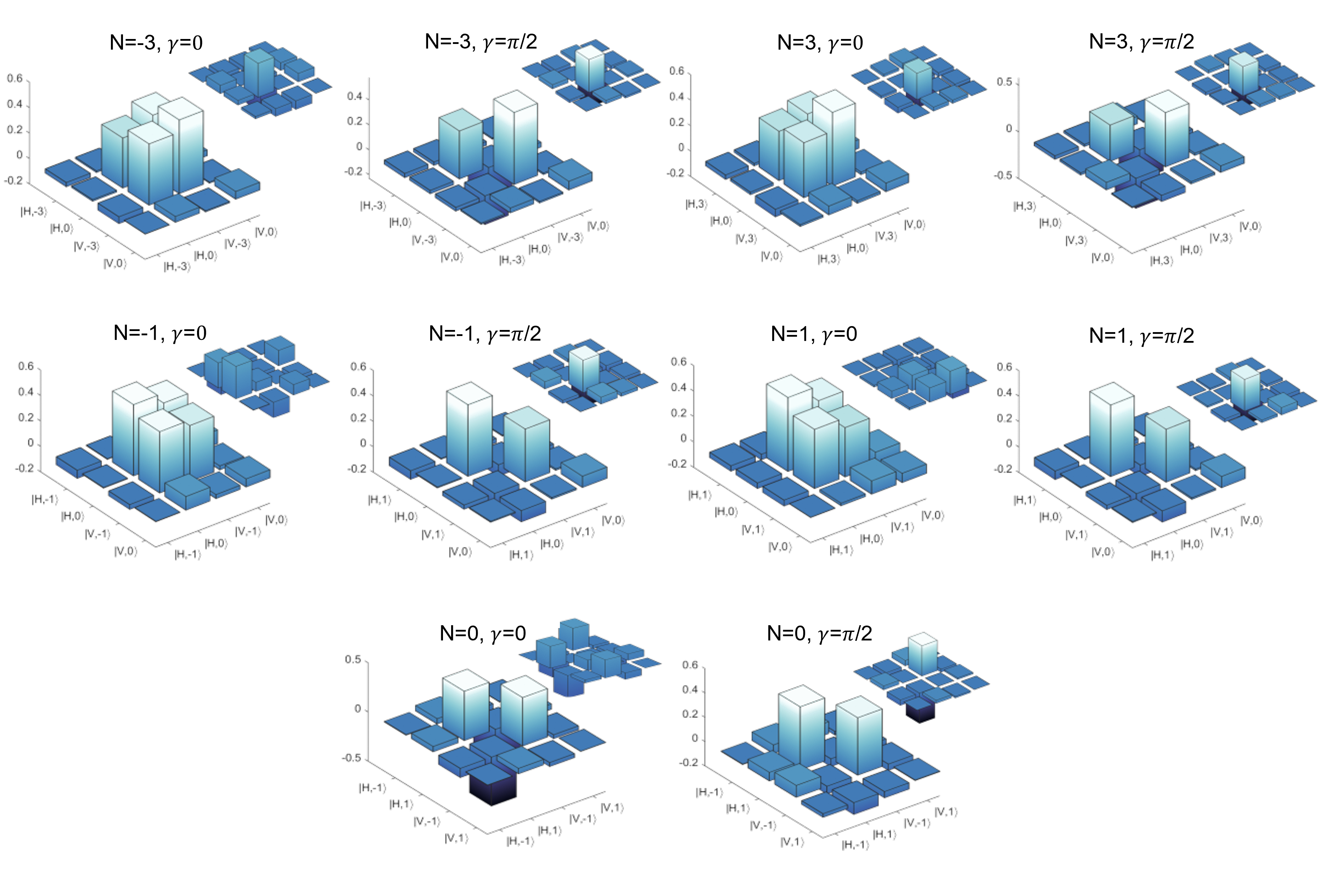}
\caption{\textbf{Reconstructed Density Matrices.} The real part of the reconstructed density matrices (imaginary parts are shown as insets) for all the data discussed in the paper is shown above. The skyrme number and $\gamma$ values for each Density Matrix representing our created states are also shown. The corresponding obtained fidelities when compared with theoretically maximally entangled states, are shown in table \ref{tbl: Data}.}
\label{fig:DensFig}
\end{figure*}

\begin{figure*}[t!]
\includegraphics[width=\textwidth]{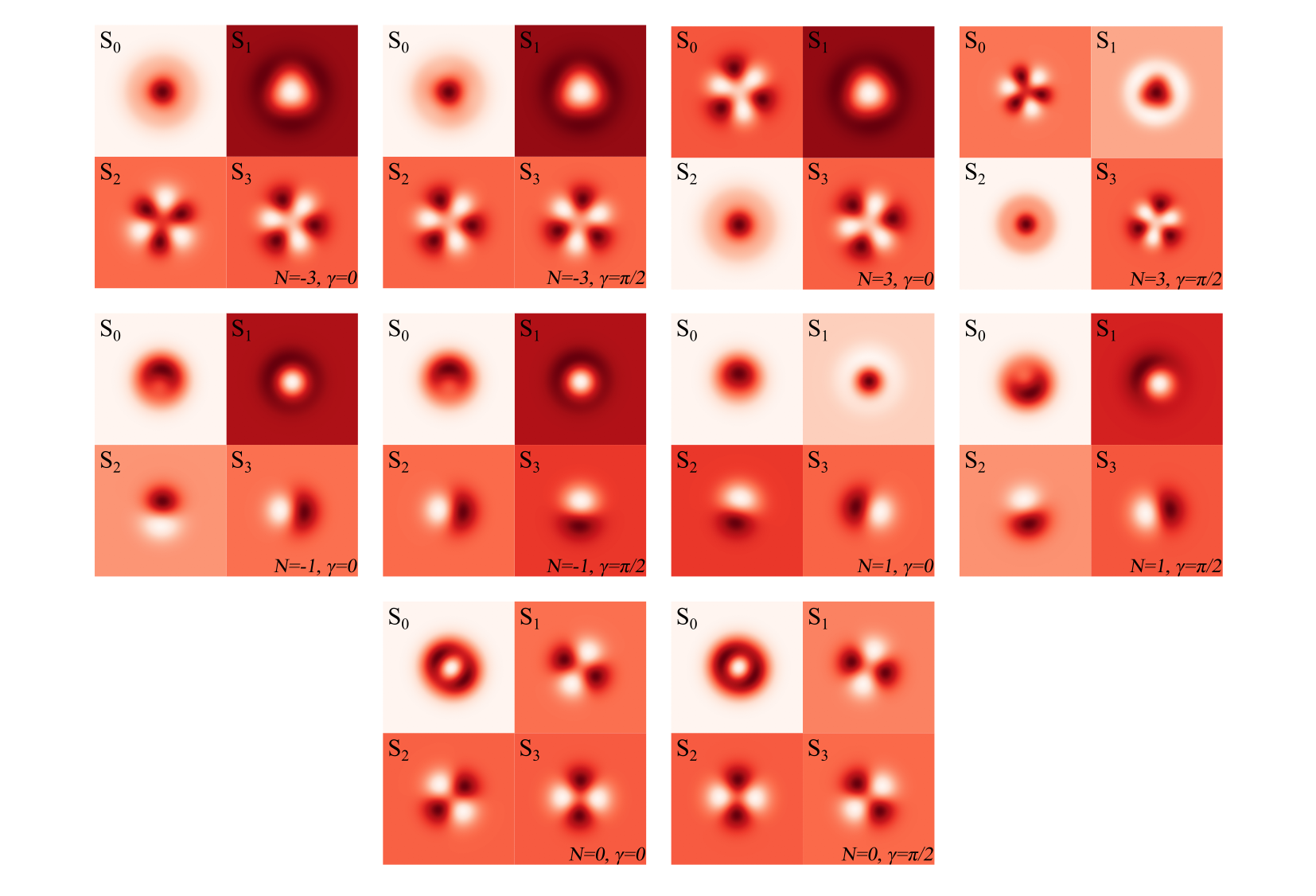}
\caption{\textbf{Reconstructed Quantum Stokes.} The reconstructed Stokes parameters for all the data discussed in the paper is shown above, with skyrme number and $\gamma$ given as insets. The data shown above is globally normalised against $S_0$, however when calculating the skyrme number, each field is locally normalized such that $S_1(x,y)^2 + S_2(x,y)^2 + S_3(x,y)^2 = 1 \forall x,y \in R^2$ in order for every point of the Stokes vector field to map onto the surface of the Poincar\"e sphere.}
\label{fig:StokesFig}
\end{figure*}

\begin{figure*}[t!]
\includegraphics[width=\textwidth]{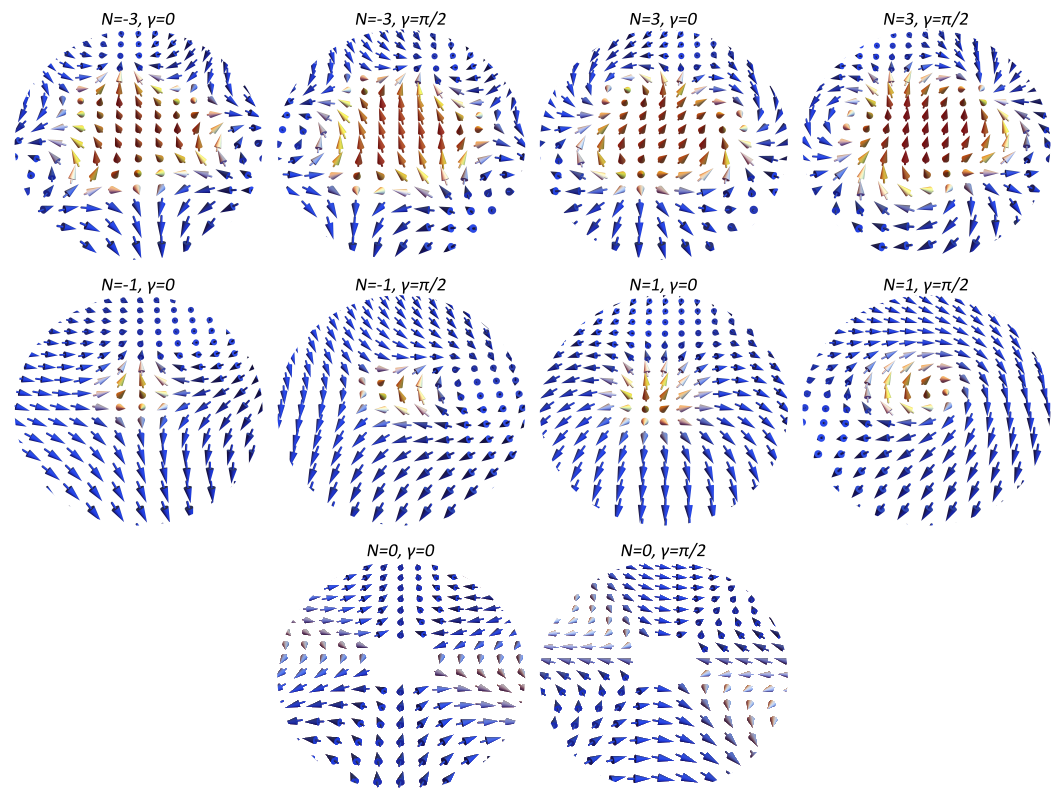}
\caption{\textbf{Reconstructed state topologies.}  The measured topological structure for all the data discussed in the paper is shown above. Vector fields for states with varying topologies (i.e varying skyrme number) and $\gamma$ are given as insets. The above vector fields are formed by sub-sampling the Stokes vector, $\Vec{S}$.  As expected, since the entangled states were created in the $|H\rangle, |V\rangle$ basis, which corresponds to the stokes parameter $S_1 = S_x$, we have that at the centre of the vector field, the vectors point to the left, corresponding to vertical polarization and as one moves away from the centre we attain elliptical polarization states (indicated with arrows pointing in and out of the plane) till at the periphery we obtain vectors pointing to the right indicating horizontal polarization.}
\label{fig:TopFig}
\end{figure*}

\end{document}